\documentclass[a4paper,twocolumn,eqsecnum,11pt,accepted=2022-06-07]{quantumarticle}
\pdfoutput=1
\usepackage[utf8]{inputenc}
\usepackage[english]{babel}
\usepackage[T1]{fontenc}
\usepackage{amsmath}
\usepackage{amssymb}
\usepackage[numbers,sort&compress]{natbib}

\usepackage{tikz}
\usepackage{lipsum}

\usepackage{tabularx}

\usepackage{xcolor}
\usepackage[
  margin=2.5cm,
  includefoot,
  footskip=20pt,
]{geometry}
\usepackage[ colorlinks = true,
             linkcolor = blue,
             urlcolor  = blue,
             citecolor = blue,
             anchorcolor = green,
]{hyperref}

\usepackage{graphicx}% Include figure files
\usepackage{dcolumn}% Align table columns on decimal point
\usepackage{bm}% bold math
\usepackage{cleveref}
\usepackage{dutchcal}
\usepackage[utf8]{inputenc}
\usepackage{nag}
\usepackage{graphicx}
\usepackage{amsthm}
\usepackage{tabulary}
\usepackage{qcircuit}
\usepackage{mathtools}
\usepackage{bm}
\usepackage{xcolor}
\usepackage{braket}
\usepackage{datetime}
\usepackage{stackrel}
\usepackage{stmaryrd}
\usepackage{dsfont}
\usepackage[english]{babel}
\usepackage{units}
\usepackage{float}
\usepackage{relsize}
\usepackage{dsfont}
\usepackage{tikz}

\newcommand{\negspace}{\!}
\newcommand{\lsub}[2]{{\protect\vphantom{#1}}_{#2} \negspace {#1}}
\newcommand{\rsub}[2]{{#1} \negspace {\protect\vphantom{#1}}_{#2}}

\newcommand{\ketsub}[2]{\rsub {\ket{#1}} {#2}}
\newcommand{\brasub}[2]{\lsub {\bra{#1}} {#2}}

\newcommand{\pbra}[1]{\brasub{#1} P}
\newcommand{\qbra}[1]{\brasub{#1} Q}
\newcommand{\pket}[1]{\ketsub{#1} P}
\newcommand{\qket}[1]{\ketsub{#1} Q}

\newcommand{\reals}[0]{\mathbb{R}}

\newcommand{\sq}{\mathcal{r}}
\newcommand{\subt}{\mathcal{S}}

\newcommand{\op}[1]{\hat{#1}}

\newcommand{\id}[0]{\op{\mathds{1}}}

\newcommand{\controlled}[1]{\op{\mathrm{C}}_{#1}}
\newcommand{\CZ}[0]{\controlled Z}

\newcommand{\opCZ}[0]{\op{\mathrm{C}}_{Z}}

\newcommand{\qwax}[1][-1]{\ar @{->} [#1,0]}
\newcommand{\bsbal}[1]{\qwax[{#1}] \qw}

\newcommand{\bsop}{\op{B} }
\newtheorem{lemma}{Lemma}

\usepackage{ulem}

\begin{document}

\title{Measurement-based generation and preservation of cat and grid states within a continuous-variable cluster state}
\author{Miller Eaton}%\textsuperscript{1}}
 \address{Department of Physics, University of Virginia, Charlottesville, VA 22904, USA}
 \address{QC82, College Park, MD 20740, USA}
\email{me3nq@virginia.edu}
\author{Carlos Gonz\'alez-Arciniegas}
 \address{Department of Physics, University of Virginia, Charlottesville, VA 22904, USA}
\author{Rafael N. Alexander}
 \address{Centre for Quantum Computation and Communication Technology, School of Science, RMIT University, Melbourne, VIC 3000, Australia}
\author{Nicolas C. Menicucci}
 \address{Centre for Quantum Computation and Communication Technology, School of Science, RMIT University, Melbourne, VIC 3000, Australia}
\author{Olivier Pfister}
 %\address{$^1$Department of Physics, University of Virginia, Charlottesville, VA 22904, USA}
 \address{Department of Physics, University of Virginia, Charlottesville, VA 22904, USA}

\date{July 11, 2022}
\begin{abstract}
We present an algorithm to reliably generate various quantum states critical to quantum error correction and universal continuous-variable (CV) quantum computing, such as Schr\"odinger cat states and Gottesman-Kitaev-Preskill (GKP) grid states, out of Gaussian CV cluster states. Our algorithm
is based on the Photon-counting-Assisted Node-Teleportation Method (PhANTM), which uses standard Gaussian information processing on the cluster state with only the addition of local photon-number-resolving measurements. We show that PhANTM can apply polynomial gates and embed cat states within the cluster. This method stabilizes cat states against Gaussian noise and perpetuates non-Gaussianity within the cluster. We show that existing protocols for breeding cat states can be embedded into cluster-state processing using PhANTM.
\end{abstract}
\maketitle
%\tableofcontents
\section{Introduction}

Quantum computation (QC), with its potential for an exponential speed-up over classical methods, is the ultimate form of quantum information processing. In contrast to traditional circuit-based QC where a register of qubits must be coherently manipulated and subjected to controlled unitary evolution~\cite{Nielsen2000}, one-way QC with cluster states allows for all entanglement to be generated up front, with single qubit measurements sufficient to enact universal QC~\cite{Raussendorf2001}. Measurement-based QC with cluster states has also been formulated for continuous-variables (CV), where now each discrete qubit can be replaced by a continuous qumode~\cite{Menicucci2006}. This is a promising area of research as CV clusters have proven advantageous over qubit-based methods in terms of massive scalability in both the frequency~\cite{Chen2014} and time domains~\cite{Yokoyama2013, Larsen2019,Asavanant2019}. These CV cluster states are formed by individually and jointly squeezed neighboring qumodes of the cluster graph. Note that the finite nature of realistic squeezing is no barrier to fault-tolerant universal QC. Indeed, when the Gottesman-Kitaev-Preskill (GKP) ``qubit in an oscillator'' encoding~\cite{Gottesman2001} is used, a fault-tolerance threshold has been shown to exist~\cite{Menicucci2014ft} for squeezing values compatible with the state of the art in optics~\cite{Vahlbruch2016}, and that threshold has been further lowered to squeezing values below 10 dB~\cite{Fukui2018}.\\
\indent However, the canonical cluster state is Gaussian in nature and although single-mode homodyne detection suffices to enact all Gaussian gates (the CV analog of the Clifford gate set)~\cite{Gu2009}, universal QC will require the addition of non-Gaussian resources~\cite{Lloyd1999,Bartlett2002,Mari2012}, such as Schr\"odinger cat or GKP states. Note that the well known Clifford-Gaussian correspondence~\cite{Lloyd1999,Gottesman2001,Bartlett2002} applies to exponential speedup in QC under the CV analog~\cite{Bartlett2002,Mari2012} of the Gottesman-Knill theorem~\cite{Gottesman1999a} but this correspondence somewhat breaks down for quantum error correction, which can be done with all-Clifford resources~\cite{Nielsen2000} but not with all-Gaussian ones~\cite{Niset2009}. Of course, employing non-Gaussian resources such as GKP states does allow quantum error correction, be it over qubits~\cite{Gottesman2001} or over purely CV qumodes~\cite{Noh2020}. Note also the remarkable result that the GKP qubit encoding allows not only quantum error correction but also universal QC with only Gaussian operations~\cite{Baragiola2019}.\\
\indent GKP states have been demonstrated experimentally with circuit QED and trapped ion systems~\cite{Fluhmann2019,CampagneIbarcq2020,de2020error}, but an all-optical approach with traveling fields has remained elusive. Many approaches have been proposed for creating and enlarging cat states and GKP states optically~\cite{Dakna1998,Ourjoumtsev2006,Vasconcelos2010, Eaton2019, thekkadath2020engineering,takase2021generation,tzitrin2020progress,Motes2017,shi2019fault,Su2019}, and several approaches have successfully generated low-amplitude cat states~\cite{Ourjoumtsev2007,Takahashi2008,Gerrits2010,etesse2015experimental, huang2015optical,ulanov2016loss, Sychev2017}. Unfortunately, all approaches currently rely on nondeterministic, probabilistic methods with success rates below $10\%$~\cite{takase2021generation} which demand ``repeat-until-success'' approaches~\cite{Knill2001,Bourassa2021blueprintscalable}. %\add{Note also a recent negative result concerning the universality of some of these methods~\cite{Gagatsos2021}.}

Last but not least, it is desirable to have these non-Gaussian states somehow integrated within the Gaussian cluster state to utilize the fully measurement-based advantages of one-way QC, which current proposals plan to do by generating non-Gaussian states offline and later coupling the ancillary modes to the cluster~\cite{takeda2019toward,Bourassa2021blueprintscalable,larsen2021fault}.  

In this article, we present a {\em near-deterministic} approach to generating non-Gaussian states, such as cat and GKP states, directly {\em embedded} in a CV cluster state, by fully leveraging the cluster state's measurement-based QC capability. This approach, which we dub the Photon-counting-assisted Node-Teleportation Method (PhANTM), allows us to build polynomial gates %based on the number of photons subtracted to
and use them to generate cat states. An essential feature of this method is that it also stabilizes cat states against Gaussian noise and prevents amplitude decay. As such, this protocol preserves non-Gaussianity in the CV cluster, which was previously thought only possible through the use of GKP error-correction~\cite{Menicucci2006}.

Practically, all PhANTM requires is an initial CV cluster state and its standard QC tools that are balanced homodyne field measurements and field displacements, to which we adjoin  photon-number-resolving (PNR) measurements in the small number regime by way of photon subtraction. 

This paper is organized as follows:  section~\ref{sec:PhANTM} describes the PhANTM algorithm and how it can be applied by performing measurements on a cluster state to generate cat states. The process is first motivated with ideal cluster states and subsequently extended to realistic cases with finite squeezing and amplitude damping introduced by experimental photon subtraction. Section~\ref{sec:results} presents numerical results demonstrating how a 1D cluster state can be reduced to a large amplitude cat state with high probability and additionally shows that the PhANTM algorithm can be used to preserve cat states already present within a cluster state. Section~\ref{sec:breeding} applies previous results on breeding cat states with beamsplitters and homodyne detection~\cite{Lund2004,oh_efficient_2018,etesse2014proposal,Vasconcelos2010,Weigand2018} to breeding cat states within the cluster state for generating enlarged cats and grid states, the later of which can be produced without post-selection~\cite{Weigand2018} and used as a resource for universal QC~\cite{Baragiola2019}. We also make the connection to phase-estimation protocols and show how they can be implemented with cat states in the cluster. Section~\ref{sec:macronode} motivates an extension to macronode-based cluster states, which are the predominant form of current CV cluster-state architectures~\cite{Yokoyama2013,Chen2014,Asavanant2019,Larsen2019}, and in Sec.~\ref{sec:conclusion} we conclude.  

It should be emphasized that this approach utilizes feed-forward Gaussian quantum information processing steps on the cluster state interspersed between PNR detection events, and as such is beyond the scope of Gaussian Boson Sampling (GBS) machines previously proposed for non-Gaussian state generation~\cite{Su2019}, in which PNR detection acts terminally. Note also the recent negative result concerning the impossibility of generating general entangled cat states with a GBS-type machine where displacements are neglected\cite{Gagatsos2021}.

\section{PhANTM}
\label{sec:PhANTM}
\subsection{Ideal Teleportation}
Measurement-based CVQC with cluster states is fundamentally based on CV teleportation, where Gaussian measurements in the form of homodyne detection teleport quantum information between neighboring nodes of a graph. The freedom to choose the measurement basis of the the homodyne detection by controlling a classical local oscillator phase additionally allows for performing Gaussian operations on the teleported state~\cite{Menicucci2006,Gu2009}. For quadrature operators defined in terms of the creation and annihilation operators as 
\begin{align}
    \op{Q}&=\tfrac{1}{\sqrt{2}}(\op{a}+\op{a}^\dag)\\
    \op{P}&=\tfrac{i}{\sqrt{2}}(\op{a}^\dag-\op{a}),
\end{align}
the ideal teleportation circuit in canonical form, read right to left, is described by 
\begin{equation}\label{eq:canon_telep}
\begin{split}
\hspace{20mm}
    \Qcircuit @C=1.5em @R=1.2em {
\lstick{\brasub{m}{P}} &\qw& \ctrl{1} & \rstick{(\text{in})} \qw \\
\lstick{(\text{out})} &\qw&\control \qw &\rstick{\ketsub{0}{P}} \qw
%&\qw&\control \qw &\rstick{\ketsub{0}{P_2}} \qw
}\raisebox{-.5em}{\hspace{8mm},}
\end{split}
\end{equation}
where $P$-subscripts indicate the quadrature basis. In this circuit, quantum information enters from the right and is coupled to a zero momentum eigenstate, $\ketsub{0}{P}$, with the control-Z gate defined as $\CZ=e^{i\op{Q}_1 \op{Q}_2}$. A homodyne measurement is then performed in the $P$ basis on the top wire yielding result $m$. We adopt the convention that circuits proceed from right-to-left, so as to coincide with the order of operator action when writing out the mathematics. The result of this circuit is simply to teleport the input quantum information in the top wire to the bottom wire with an additional Fourier transform (rotation of $\pi/2$) and a displacement that depends on the measurement result; that is to say, the operator $\op{X}(m)\op{R}(\tfrac{\pi}{2})$ is applied to the input. Here, we have defined the rotation and quadrature shift operators as
\begin{align}
    \op{Z}(y)&=e^{i\op{Q}y}\\
    \op{X}(x)&=e^{-i\op{P}x}\\
    \op{R}(\theta)&=e^{i\theta \op{a}^\dag \op{a}},
\end{align}
where a general displacement is expressed as
\begin{align}
    \op{D}(\alpha)&=e^{\alpha \op{a}^\dag - \alpha^{\!*} \op{a}}\nonumber \\
    &=\scriptstyle{\op{Z}\left(\sqrt{2}\text{Im}[\alpha]\right)\op{X}\left(\sqrt{2}\text{Re}[\alpha]\right)}\normalsize{e^{-i\text{Re}[\alpha]\text{Im}[\alpha]}}.
\end{align}
By varying the measurement phase, all Gaussian operations on cluster states can be realized with homodyne measurement alone\cite{Gu2009} up to a displacement. 

However, in order to achieve a quantum speed-up, non-Gaussianity must be introduced; and furthermore, this non-Gaussianity must not be efficiently simulable classically~\cite{chabaud2021classical}. This can be done by introducing the ability to perform photon addition or subtraction, which introduces negativity into the multi-mode Wigner function~\cite{Walschaers2018,walschaers2020practical}. In our approach, we examine the effects of applying PhANTM, which utilizes PNR detection to perform photon-subtraction within the teleportation circuit. 

Suppose we wish to perform a photon subtraction of $n$ photons immediately prior to the homodyne detection step of the teleportation circuit. This circuit, including finite squeezing on the ancillary mode, is represented by
\begin{align}\label{eq:phot_sub_circ}
    \begin{split}
    \Qcircuit @C=1.5em @R=1.5em 
    {
	&\lstick{\pbra{m}}  & \qw   & \gate{\subt_n} &\ctrl{1} &\qw& \rstick{(\text{in})} \qw \\
	&\lstick{\text{(out)}}	&\qw & \qw &\control \qw &\gate{S(\sq)}&  \rstick{ \ketsub{0}{N} } \qw \\
		}\, 
	\end{split} \hspace{8mm}.	
\end{align}
Note that we use the calligraphic r, $``\sq"$, for the squeezing parameter so as not to confuse with $r$ used for beamsplitter reflectivity. In this circuit, a realistic photon subtraction of $n$ photons is implemented by applying the Kraus operator
\begin{equation}
\op{\mathcal{S}}_n=\frac{(-1)^ne^{n\beta/2}}{\sqrt{n!}}\left(2\sinh{\beta}\right)^{n/2}\hat{a}^n \hat{N}(\beta),
    \label{eq:kraus_photsubt}
\end{equation}
where
\begin{equation}
    \hat{N}(\beta):=e^{-\beta \hat{a}^\dag \hat{a}}=e^{-\tfrac\beta2(\op{Q}^2+ \op{P}^2 - 1)}
\end{equation}
is the damping operator with $\beta \in \reals$. Here, a squeezed momentum state is achieved by applying the squeezing operator
\begin{equation}
    S(\sq)=e^{\tfrac{\sq}{2}({\hat{a}^{\dag 2}}-\op{a}^2)}
    \end{equation}
to the vacuum state, denoted as a zero-valued ket with subscript $N$ denoting the photon-number eigenbasis. Eq.~\ref{eq:kraus_photsubt} can be implemented by coupling the top wire to a vacuum mode with a beamsplitter of transmissivity $t=e^{-\beta}$ and then performing PNR detection on the reflected mode as derived in Appx.~\ref{Appx:phot_subt}. Additionally, Appx.~\ref{appx:damp} gives further information about the non-unitary damping operator. \\
\indent Before analyzing this circuit in depth, we wish to simplify it to a mathematically idealized case where its usefulness is more readily apparent.
First, we take the high squeezing limit such that we can approximate the squeezed vacuum as a zero momentum eigenstate, $\op{S}(\sq)\ketsub{0}{N}\approx\pket{0}$. Next, consider taking the limit $\beta\rightarrow 0$ in Eq.~\ref{eq:kraus_photsubt}, which is physically equivalent to tuning the subtraction beamsplitter to have vanishing reflectivity. Measuring $n$ photons in the weakly reflected mode will result in the application of $\subt_n\approx \op{a}^n$, up to a normalization. In this limit, the probability to measure photon numbers with $n>0$ vanishes; however, when properly combined with the high squeezing limit such that the beamsplitter reflectivity becomes small as the energy in the input state becomes very large, the probability to subtract zero photons can also vanish. Later, we will relax both constraints on the squeezing and beamsplitter reflectivity (value of $\beta$) to demonstrate that our method can perform successfully in a realistic scenario. 

For now, consider the idealized non-Gaussian subtraction and teleportation circuit given by
\begin{equation}\label{eq:canon_subtr}
    \Qcircuit @C=1em @R=1.5em {
\lstick{\brasub{m}{P}} &\qw&\gate{a^n} &\ctrl{1} & \rstick{(\text{in})} \qw \\
\lstick{(\text{out})} &\qw&\qw&\control \qw &\rstick{\pket{0}} \qw
%&\qw&\qw&\control \qw &\rstick{\pket{0}} \qw
}\raisebox{-1em}{\hspace{10mm}.}
\end{equation}
Writing the annihilation operator as $\hat{a}=\tfrac{1}{\sqrt{2}}(\op{Q}+i\op{P})$ and commuting this with the $\opCZ$ gate, we have that the operator applied to the output state is given by
\begin{align}
   &\brasub{m}{1P}\frac{\opCZ}{\sqrt{2^{n}}}\left(\op{Q}_1+i\op{P}_1+i\op{Q}_2\right)^{n}\ketsub{0}{2P}\nonumber \\
    =&\brasub{m}{1P}\frac{\opCZ}{\sqrt{2^{n}}}\sum_{k=0}\binom{n}{k}\left(i\op{Q}_2\right)^{k} (\op{Q}_1+i \op{P}_1)^{n-k}\ketsub{0}{2P}.
\end{align}
Since the $\opCZ$ gate commutes with $\op{Q}$, we can graphically represent each term in the sum as the circuit
\begin{align}
\centering
\hspace{8mm}
\begin{split}
        \Qcircuit @C=0.8em @R=1em {
\lstick{\brasub{m}{P}} &\qw&\qw &\ctrl{1} &\gate{(Q+iP)^{n-k}}& \rstick{(\text{in})} \qw \\
\lstick{(\text{out})} &\qw&\gate{(iQ)^{k}}&\control \qw &\qw&\rstick{\pket{0}} \qw
% &\qw&\gate{(iQ)^{k}}&\control \qw &\qw&\rstick{\pket{0}} \qw
}\raisebox{-1.5em}{\hspace{10mm},}
\end{split}
\end{align}
which is simply the teleportation circuit applied to a modified input followed by the application of $\op{Q}$ to a power. Representing the action of this circuit as a Kraus operator applied to the input, we see that the resultant operator would be
\scalebox{0.9}{
\begin{math}
\begin{aligned}
\label{eq:SAT_ideal}
&\op{K}_{m,n}=\frac{1}{\sqrt{2^{n}}} \sum^n_{k=0}\binom{n}{k}\left(i\op{Q}\right)^{k} \op{R}(\tfrac\pi2)\op{Z}^\dag(m)(\op{Q}+i \op{P})^{n-k} \nonumber\\
&=\op{X}(m)\op{R}(\tfrac\pi2)\frac{1}{\sqrt{2^{n}}} \sum^n_{k=0}\binom{n}{k}\left(-i\op{P}+im\right)^{k} (\op{Q}+i \op{P})^{n-k}.
\end{aligned}
\end{math}

}
The position shift of $\op{X}(m)$ at the end can be effectively ignored, as its effect can be removed with feed-forward displacements or accounted for by shifting the result of subsequent homodyne detections~\cite{Gu2009}. Surprisingly, the $\op{P}$ contribution in the sum vanishes and the operator can be written as $\op{K}_{m,n}=\op{X}\op{R}f_n(\op{Q})$, where $f_n(\op{Q})$ takes the form of a polynomial with generally complex coefficients. This polynomial has implicit $m$ dependence and will be given in the next section. This can be seen by examining the commutator of $\op{K}_n$ with $\op{Q}$. Furthermore, in the specific case where $m=0$, Eq.~\ref{eq:SAT_ideal} can be written in terms of a Hermite polynomial in $\op{Q}$ as
\begin{equation}
    \op{K}_{0,n}=\op{R}(\tfrac{\pi}{2})H_n({\tfrac{\op{Q}}{\sqrt{2}})}
    \label{eq:SAT_ideal_hermite},
\end{equation}
where $H_n(x)$ is the $n$-th order physicist's Hermite polynomial. This is derived in Appx.~\ref{Appx:herm_proof}. By iterating this procedure $M$ times successively, actively undoing the displacements between steps, and applying a regular teleportation at each intervening step to enact a Fourier transform and keep the overall operator a function of $\op{Q}$ only, we can build up powers of $\op{Q}$ to form the overall operator
\begin{equation}
    \op{K}_M=\op{R}(M\pi)\prod_{k}f_k\left((-1)^k\op{Q}\right),
    \label{eq:gen_qop}
\end{equation}
which is a polynomial in $\op{Q}$ follow by a rotation of $M\pi$. The degree of this polynomial is equal to the total number of photons subtracted over all steps. It is worth pointing out that while we have presented this analysis from the standpoint of the idealized circuit in Eq.~\ref{eq:canon_subtr}, including realistic effects will still result in the application of a polynomial in $\op{Q}$ of the same order, but with additional Gaussian noise. Before continuing on to the full treatment, we wish to further motivate the use of applying polynomials of this form. Developing polynomial gates can be used to approximately implement non-Gaussian unitaries and the cubic phase gate~\cite{Marshall2015,arzani2017polynomial}, but when specifically applied to squeezed vacuum, they can give rise to cat states~\cite{Ourjoumtsev2007}, which is our main interest here.

\subsubsection{Polynomial Operator Applications}
\begin{figure}[ht]
 \includegraphics[width = 0.45\textwidth]{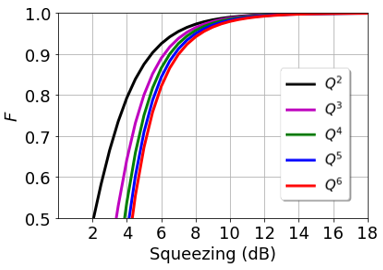}
  \caption{Application of Hermite polynomials in $\op{Q}$ compared to leading order powers of $\op{Q}$. Fidelity approaches unity as squeezing increases.}
    \label{fig:QvsHerm}
\end{figure}

As a segue to a fully realistic implementation, consider applying this idealized overall $\op{K}_M$ operator to a cluster-state node with finite squeezing, and consider the case where all homodyne measurements are null-valued (a constraint we will relax later), in which case $f_n(\op{Q})$ is a Hermite polynomial. For moderate to large squeezing on the state to which the polynomial operator is applied, the leading power of $\op{Q}$ in each Hermite polynomial will dominate as the anti-squeezing gives support over a large range of position in phase-space, thus eliminating the contribution of lower order terms upon normalizing the state after the operator is applied. This is demonstrated in Fig.~\ref{fig:QvsHerm}, where the fidelity of $H_n(\op{Q}/\sqrt{2})S(\sq)\ketsub{0}{N}$ and $\op{Q}^n S(\sq)\ketsub{0}{N}$ is plotted as a function of squeezing. We use the fidelity between two density matrices $\rho$ and $\sigma$ defined as
\begin{equation}
\label{eq:fid_def}
    F =\text{Tr}\left[\sqrt{\sqrt{\sigma}\rho\sqrt{\sigma}}\right]^2.
\end{equation}For $n=1$, the cases are trivially the same, and for larger $n$, the fidelity approaches one as squeezing increases. Thus, for the immediate sake of illustration, we can examine the effects of applying only $\op{Q}^M$ to a squeezed state, where $M$ is now the total number of photons subtracted over all iterations of the process.

\begin{figure*}[ht]
\centering
\includegraphics[width = 0.85\textwidth]{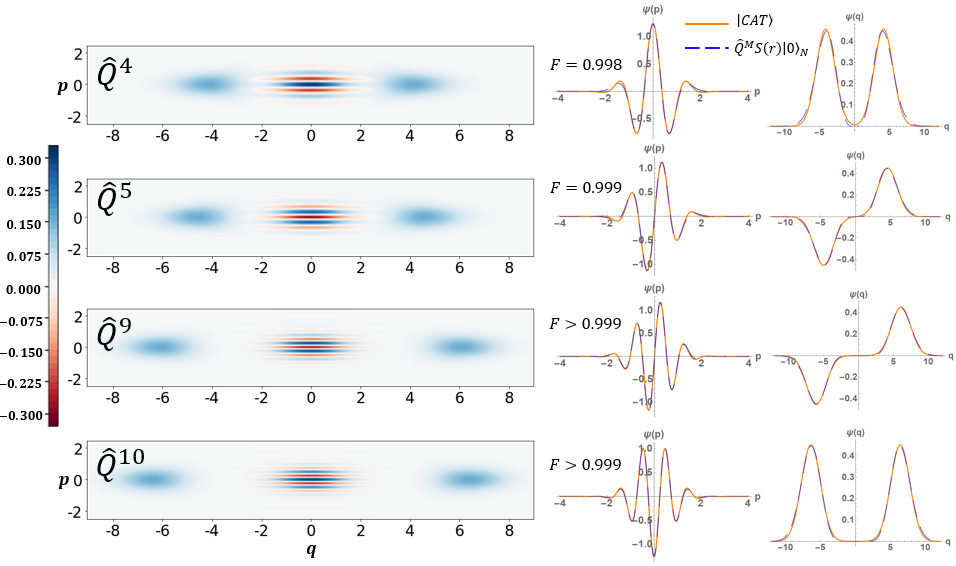}
  \caption{Idealized PhANTM steps with null homodyne measurement results iteratively apply powers of $\op{Q}$ to an input state based on the total number of subtracted photons over all rounds of teleportation. Here, the input is a squeezed vacuum state with squeezing of 6 dB ($r=0.7$). The final state is a nearly perfect match to an even (odd) squeezed cat state when an even (odd) number of photons was subtracted.  }
    \label{fig:cat_wig}
\end{figure*}

The Wigner functions of the resulting states after applying $\op{Q}^M$ to squeezed vacuum for several values of $M$ are plotted in Fig.~\ref{fig:cat_wig}. As one can see the results resemble anti-squeezed Schr\"odinger Cat states, where increasing the power of $Q$ applied to the squeezed state increases the amplitude of the approximated cat. The closeness to cat states can be seen by looking at the form of the wavefunction. In the $Q-$basis, the state wavefunction is given by
\begin{equation}
\label{eq:approx_cat_psi}
   \psi_Q(v)= \qbra{v} Q^M S(\sq)\ketsub{0}{N}=\mathcal{N}e^{-(\frac{v^2}{2s^2})}v^M,
\end{equation}
where $s=e^{\sq}$ and the normalization is given by
\begin{equation}
    \mathcal{N}=s^{-M}\sqrt{\frac{1}{s(M-\tfrac{1}{2})!}}.
\end{equation}

Transforming to the $P-$basis, we have 
\begin{equation}
    \psi_P(t)=\mathcal{N'}e^{(-\tfrac{s^2t^2}{2})}H_M(\tfrac{st}{\sqrt{2}}),
\end{equation}
where $\mathcal{N'}=(-2)^{-\tfrac M2}s^{M+\tfrac12}\mathcal{N}$ and $H_M(x)$ are Hermite polynomials. This wavefunction resembles that of a squeezed $M-$photon Fock state, but with the subtle difference that the argument of the Hermite polynomial is scaled by a factor of $2^{-1/2}$. No additional squeezing operation will transform it back to a Fock state as scaling the argument of the exponent will also scale the argument of the Hermite polynomial. Furthermore, this scaling is independent of the power of $Q$ applied. Examining this wavefunction shows that the number of ripples is determined by the order of the Hermite polynomial, which is based on the total photon subtraction counts. As shown in Fig.~\ref{fig:cat_wig}, a nearly identical anti-squeezed cat state can be found for each of these, where fidelities with cats having numerically optimized parameters are above $0.999$. Additionally, it has previously been shown that the wavefunction in Eq.~\ref{eq:approx_cat_psi} asymptotically approaches a cat state as $M$ becomes large~\cite{Ourjoumtsev2007}.

\begin{figure}[!htb]
 \includegraphics[width = 0.5\textwidth]{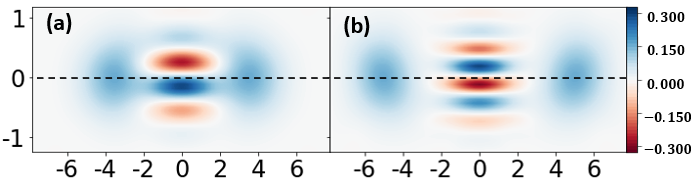}
  \caption{Typical results from iterated subtracted teleportation applied to squeezed vacuum with one photon subtracted each instance and stochastic homodyne results. The dotted black line is added as a guide to show the phase of the resultant cat state. Note that the figures have been rescaled to better visualize the fringes.}
    \label{fig:stoch_HD}
    \end{figure}

As a final step before proceeding to the full treatment, we relax the restrictions imposed on homodyne measurement and allow for arbitrary quadrature detection results. This leads us to the slightly more general operator given by a product of Hermite polynomials in $\op{Q}$ with arguments shifted by the measurement result, which arises from a limiting case of the general derivation that will be discussed later (Eq.~\ref{eq:limit_case_final_appx} in Appx.~\ref{Appx:main_derivation}). However, random homodyne measurement results do not pose a significant obstacle for cat-state generation. This effect leads to producing cat states with a general phase between the `classical' displacement components of the form:
\begin{equation}
    \ket{cat}=\mathcal{N_\alpha}\left(D(\alpha)+e^{i\phi}D(-\alpha)\right)S(\sq)\ketsub{0}{N}.
    \label{eq:cat}
\end{equation}
As shown in Fig.~\ref{fig:stoch_HD}, protocols with stochastic homodyne detection generate cat states with a phase-space interference fringe that has been scanned in a direction perpendicular to the displacements. The Wigner functions plotted in Fig.~\ref{fig:stoch_HD} depict typical results of a multistage process where finite squeezing of $6$ dB has been used for all ancillary inputs, and a single photon was subtracted in each teleportation iteration before a stochastic homodyne measurement. 

\subsection{Realistic Teleportation}
Until now, we have only described an idealized case where teleportation proceeded with infinite squeezing and photon-number subtraction was modeled as a perfect application of the annihilation operator. The true circuit that represents realistic photon-subtraction during a cluster-state teleportation will instead appear as 
\begin{equation}\label{eq:subtr_tele}
    \Qcircuit @C=0.8em @R=1.5em {
\lstick{\brasub{n}{N}} &\bsbal{1} & \qw&\qw& \rstick{\ketsub{0}{N}} \qw \\
\lstick{\brasub{m}{P}}  &\qw &\ctrl{1} & \qw& \rstick{(\text{in})} \qw \\
\lstick{(\text{out})} &\qw&\control \qw & \gate{S(\sq')} &\rstick{\ketsub{0}{N}} \qw
%&\qw&\control \qw & \gate{S(\sq')} &\rstick{\ketsub{0}{N}} \qw
}\raisebox{-1.9em}{\hspace{10mm},}
\end{equation}
where the application of the annihilation operator is instead replaced by an additional circuit component to couple the input state in the middle wire to vacuum in the top wire with a beamsplitter, 
\begin{equation}
    \bsop=e^{\theta(\op{a}_1 \op{a}_2^\dag-\op{a}_1^\dag\op{a}_2)},
\end{equation} which is represented by the down arrow in the diagram. We take the beamsplitter to have real reflectivity and transmissivity such that $r=\sin{\theta}$ and $t=\cos{\theta}$. The momentum eigenstate in the third wire has also been replaced with a finitely squeezed vacuum state. In the limit of weak beamsplitter reflectivity, projecting the top wire into an $n$-photon Fock state after the beamsplitter effectively acts to apply $\op{a}^n$, although the probability to successfully measure $n$ photons vanishes as the reflectivity drops to zero. Away from this limit, this portion of the circuit acts to apply non-unitary damping to the state in addition to a factor of $\op{a}^n$, as discussed in Appx.~\ref{Appx:phot_subt}.  

The top portion of the circuit in Eq.~\ref{eq:subtr_tele} can be simplified to
\begin{equation}
\label{eq:sub2fQ}
       \begin{split}
 \Qcircuit @C=0.3cm @R=0.8cm {
     \lstick{\brasub{n}{N}} &\bsbal{1} &\qw& \rstick{\ketsub{0}{N}}&&& \\
     \lstick{\brasub{m}{P}}&\qw& \qw &&}{\raisebox{-1.3em}{$=$}}\hspace{1em}
     \Qcircuit @C=0.2cm @R=0.1cm {\vspace{2em}&&&&\\&&&&\lstick{\brasub{0}{P}}&\gate{f_n(Q)}&\qw
    }\hspace{4mm}{\raisebox{-.8em}{,}} 
       \end{split}\, 
\end{equation}\\
where $f$ is a function in $\op{Q}$ only. On the bottom wire, we can write the input momentum-squeezed state as
\begin{align}\label{eq:sq_trick}
    \hat{S}(\sq')\ketsub{0}{N}&=\pi^{-1/4}\int dt  e^{-t^2/2s^2}\qket{t}\nonumber \\
    &=\pi^{-1/4}e^{-\op{Q}^2/2s^2}\int dt \qket{t} \nonumber \\
    &=\pi^{1/4}\sqrt{\frac{2}{s}}e^{-\op{Q}^2/2s^2}\pket{0},
\end{align}
where $s=e^{\sq'}$. This allows us to commute all operations in $\op{Q}$ to either the back or front of the circuit to arrive at
\begin{equation}
    \Qcircuit @C=0.8em @R=1.2em {
\lstick{\brasub{0}{P}} &\qw&\qw&\ctrl{1} & \gate{f_n(Q)} & \rstick{(\text{in})} \qw \\
\lstick{(\text{out})} &\qw&\gate{e^{-Q^2/2s^2}}&\control \qw & \qw &\rstick{\pket{0}} \qw
%&\qw&\gate{e^{-Q^2/2s^2}}&\control \qw & \qw &\rstick{\pket{0}} \qw
}\raisebox{-1.5em}{\hspace{10mm},}
\end{equation}
which is recognizable as the same form as Eq.~\ref{eq:canon_telep} with additional operators before and after the teleportation. We can now directly write down the Kraus operator representation of the action of the circuit on the input state as
\begin{equation}
\label{eq:kraus_main}
    \op{K}_n=\pi^{1/4}\sqrt{\frac{2}{s}}e^{-\op{Q}^2/2s^2}\op{R}(\tfrac{\pi}{2})f_n(\op{Q}).
\end{equation}

The function $f_n(\op{Q})$ contains a displacement from the measurement result that may be useful to keep track of separately, so $f$ can be decomposed into a displacement term, a quadrature damping term, and a polynomial in $\op{Q}$. Commuting the displacement term through the Fourier transform allows us to arrive at 
\begin{align}
    \op{K}_n=&\tfrac{\pi^{1/4}\sqrt{2}}{\sqrt{s}}e^{\tfrac{m^2}{s^2}\sigma^2}\op{X}(m\sigma)\op{R}(\tfrac{\pi}{2})\nonumber \\
    &\times e^{-\tfrac{1}{2s^2}(\op{P}-m\sigma)^2}e^{-\op{Q}^2 \tfrac{\sigma t(1-t^2)}{4}}f_n'(\op{Q}),
    \label{eq:PhANTM_kraus}
\end{align}
where we define $\sigma=\tfrac{2t}{1+t^2}$. This is derived in detail in Appx.~\ref{Appx:main_derivation}, where we find that
\begin{widetext}
\begin{equation}
    f_n'(x)=\tfrac{i^n}{\sqrt{2^{n-1}n!(1+t^2)^{n+1}}}\sum_{k=0}^n\sum^{\left\lfloor \text{$\tfrac{n-k}{2}$} \right\rfloor}_{j=0}\binom{n}{k}\binom{n-k}{2j}\frac{(2j)!r^{2j}}{2^j j!}\left(\tfrac{1+t^2}{2}\right)^{k/2}(-i)^{k+2j}H_k(\tfrac{-x\sqrt{2}rt^2}{1+t^2})H_{{n-k-2j}}(\tfrac{-mr}{t\sqrt{1+t^2}})
   \label{eq:fprime}
\end{equation}
\end{widetext}
is a polynomial of degree $n$. In the limit of weak beamsplitter reflectivity and large squeezing so that $r\rightarrow0$, $t\rightarrow1$ and $s\rightarrow \infty $, the Kraus operator reduces to  
\begin{equation}
    \op{K}_n\propto \op{X}(m)\op{R}(\tfrac{\pi}{2})H_n\left(\frac{i\op{Q}-m}{\sqrt{2}}\right).
    \label{eq:limit_case_final}
\end{equation}
Subtracting photons during the teleportation process is stochastic in nature, where the probability of any particular $n$-photon subtraction and $m$ homodyne measurement occurring is given by
\begin{equation}
    P_n=\text{Tr}[\op{K}_n\rho_{in}\op{K}_n^\dag],
\end{equation}
where $\rho_{in}$ is the general quantum state being sent through the PhANTM gadget, and the final state is evolved to
\begin{equation}
    \rho_{out}=\frac{\op{K}_n \rho_{in}\op{K}_n^\dag}{P_n}.
\end{equation}

As in the idealized case, we assume that one has the ability to undo the accumulated displacement operations. If the photon subtraction was successful, then any Gaussian input has been de-Gaussified. If, however, no photons were subtracted, the state will accrue additional Gaussian noise due to the finite squeezing in the cluster state and the non-vanishing beamsplitter reflectivity used to perform the subtraction. For large enough squeezing and low beamsplitter reflectivity, this additional Gaussian noise can be made small. 

Whether subtraction was successful or not, teleporting the resultant state through the next node of the quantum wire without attempting to subtract will enact another $\pi/2$ rotation and re-align the squeezed-axis of the state, at which point another PhANTM step can be performed. Provided there is sufficient squeezing in the cluster state, this process of subtracting photons through teleportation can build up non-Gaussianity faster than the Gaussian noise at each step can degrade the state and result in large, high-fidelity cat states.
\begin{figure}[!htb]
 \includegraphics[width = 0.49\textwidth]{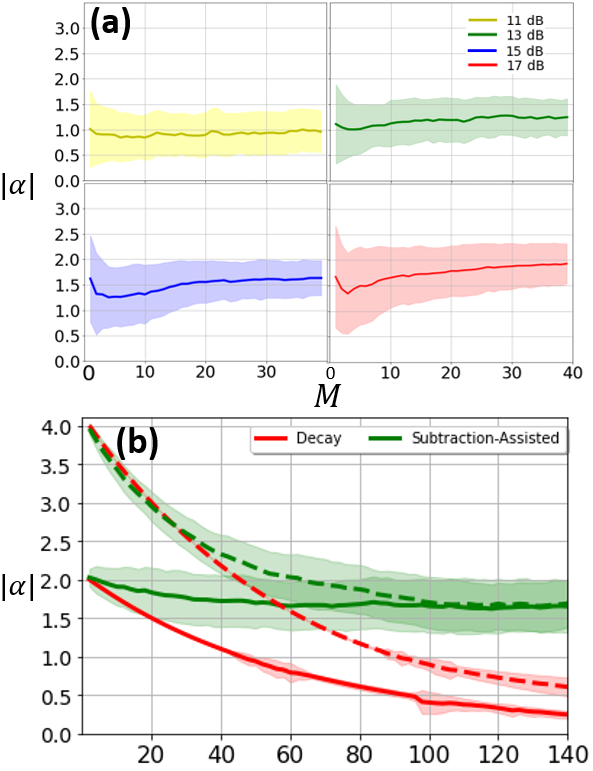}
  \caption{(a) Solid lines show the average cat amplitude obtained from numerical simulations of performing $M$ iterations of the subtraction protocol on a 1D cluster state where the homodyne and PNRD measurements were randomly sampled from the state distribution at each step. The shaded regions show $\pm\sigma$ from the sampled data and squeezing values indicate the squeezing level of each node in the initially prepared cluster state. The average fidelity with the nearest cat is above 0.98 for $M\geq10$ in all plots. (b) Cat states with respective amplitudes of $|\alpha|=2$ and $|\alpha|=4$ introduced to a cluster state can be protected as they are teleported through the cluster if subtraction is attempted every other teleportation step to prevent state decay. Shading shows $\pm\sigma$.}
    \label{fig:cat_results}
    \end{figure}
\section{Results}
\label{sec:results}
\subsection{Cat States}
\label{subsec:catstates}

In this section, we present the results of numerical calculations demonstrating that iteratively performing the PhANTM algorithm on cluster states can generate and stabilize large cat states. Even when considering finite squeezing effects and the probabilistic nature of the Kraus operator given by Eq.~\ref{eq:PhANTM_kraus}, repeatedly applying operators of this form gradually shapes the teleported quantum information into cat states and preserves cat states already present in the cluster.

For these calculations, the python packages of QuTip~\cite{Johansson2012,Johansson2013QuTiP2A} and Strawberry Fields~\cite{killoran2019strawberry,bromley2020applications} were used to perform simulations of quantum state evolution using the Rivanna high-performance computing system at the University of Virginia. A single step of the PhANTM along a 1D cluster state was first simulated by sampling the probability distribution for detection outcomes from the circuit shown in Eq.~\ref{eq:subtr_tele}, and then the input state was updated accordingly by applying the corresponding Kraus operator from Eq.~\ref{eq:kraus_main}. This protocol was applied in succession, with a regular teleportation to reapply a $\pi/2$ rotation occurring between PhANTM steps to simulate the algorithm along several nodes of a cluster state. Cluster states with different squeezing parameters were examined, and statistics were collected by simulating the entire procedure repeatedly. It should be noted that whenever a Wigner function is displayed, it is actually the single-mode Wigner function of a particular cluster-state node as if it had not yet been connected by a $\opCZ$ gate. All states are in reality embedded \textit{within} the cluster and neighboring nodes would need to be measured out to physically disconnect the particular mode in question.  

In each case, the value of beamsplitter reflectivity for the photon subtraction step was chosen so that the added Gaussian noise from subtraction effectively adds the same amount of noise as finite squeezing in the cluster. The condition is given by
\begin{equation}
    \label{eq:damp_sq_equiv}
    \cos^2{\theta}=\tanh{|\sq'|}
\end{equation}
where $\sq'$ is the squeezing parameter for the cluster-state nodes. This condition arises by realizing that applying the damping operator, $\op{N}(\beta)$, to a zero-momentum eigenstate is equivalent to finitely squeezing the vacuum state, as has been derived previously~\cite{walshe2020continuous} and discussed further in Appx.~\ref{appx:damp}. Because squeezing can instead be thought of as applying the damping operator to a momentum eigenstate and the damping operator is also what separates the realistic photon subtraction operator $\op{\mathcal{S}}_n$ from ideally applying annihilation operators, implementing the above condition essentially makes the real circuit in Eq.~\ref{eq:phot_sub_circ} different from the idealized circuit in Eq.~\ref{eq:canon_subtr} only by the inclusion of damping operators applied both to the ancilla and before homodyne detection. %{\color{red} [RNA: explain why this condition is what makes the noise equal. ]} 

For each trial, we started with a fresh 1-D cluster state and applied many steps of the PhANTM along the cluster while allowing stochastic measurement results. After each step in the simulation, the evolved state was fitted to the nearest cat state of the form of Eq.~\ref{eq:cat} by optimizing the fidelity. Fidelity and fit values for all trials with a given set of initial cluster-state squeezing parameters were averaged at each step, $M$, and compared to trials with different squeezing parameters. 

The results are depicted in Fig.~\ref{fig:cat_results}(a) for cluster states with $11-17$ dB of squeezing at each node. The average fitted value of the nearest cat-state amplitude (solid lines), $|\alpha|$, is plotted against the number of times that a PhANTM had occurred, where each step corresponds to consuming two physical nodes of the cluster due to the necessary intervening teleportation step between PhANTMs to apply a Fourier transform. After $M=10$ steps, the average fidelities with the associated cats are greater than $0.98$ in all cases examined. Increasing $M$ beyond 10 steps has little impact on the fidelity other than to increase cat amplitude, which is correlated with fidelity as mentioned earlier~\cite{Ourjoumtsev2007}. The shaded regions indicate the standard deviation in the spread of obtained values from 100 trials. 

After just a few steps, it is possible to generate a strongly anti-squeezed cat state of weak amplitude, but this is highly probabilistic and dependent on the precise subtraction measurements. However, if there is sufficient squeezing in the cluster state, each PhANTM on average succeeds in adding more non-Gaussianity than is washed away by Gaussian noise, and thus there is a high probability of obtaining a cat state with reasonable amplitude after several measurement steps. For 17dB of cluster-state squeezing, 30 steps of PhANTM are sufficient to succeed in generating a high-fidelity cat state with amplitude $|\alpha|>1.5$ with a $68\%$ success rate and $|\alpha|>1$ with a $95\%$ success rate.  It should be noted that although anti-squeezed cats are obtained in the idealized case of Fig.~\ref{fig:cat_wig}, the effects of finite squeezing in the cluster and overall stochastic nature of the photon subtraction result in cat states very low anti-squeezing ($<3$ dB) beyond the first few steps, which is the region of interest. The Wigner functions and recorded PNR subtractions for a typical example are shown in Appx.~\ref{Appx:example}.   

\subsubsection{Stabilization of Cat States}
\label{subsec:cat_stabilize}
In addition to generating cat states from a Gaussian resource using only local measurements, the PhANTM protocol can also be used to protect cat states already present within the cluster state. Suppose one has a cat state embedded in a cluster state, either generated through the PhANTM as detailed above or offline through some other means, and entangles it to a cluster state with a $\opCZ$ interaction for later use. If this state needs to be moved through the cluster to a suitable location, it will suffer amplitude decay from Gaussian noise at each teleportation step, as shown by the red lines in Fig.~\ref{fig:cat_results}(b). If one applies the PhANTM algorithm in place of regular teleportation at every-other node, then the cat state can instead be preserved (green curves) to a level dependent on the squeezing present in the cluster.

\begin{figure}[h]
 \includegraphics[width = 0.48\textwidth]{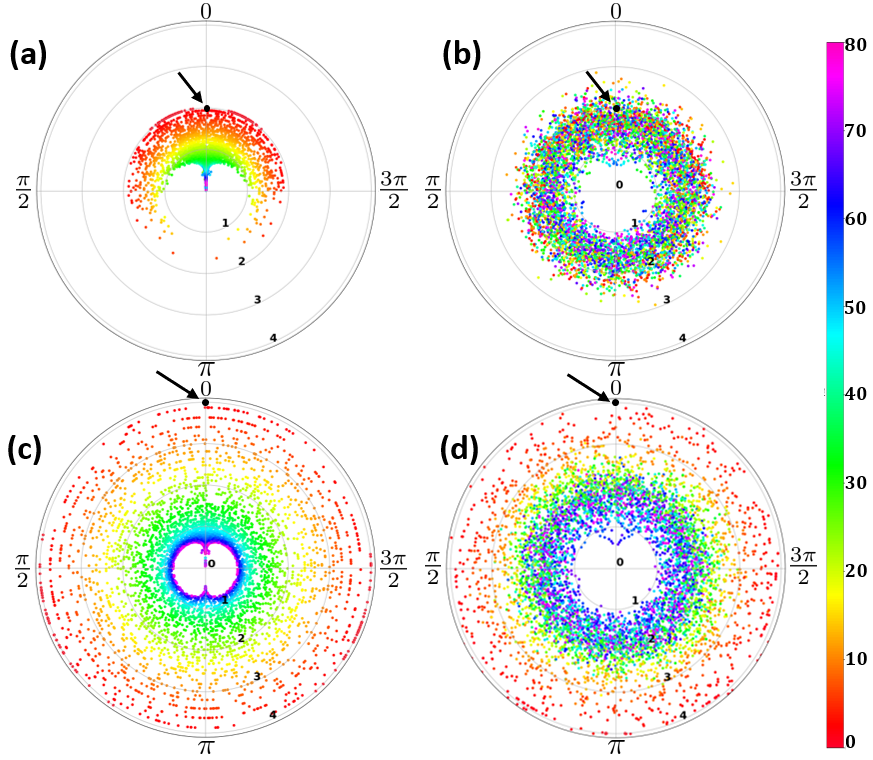}
  \caption{Alternate representation of the data in Fig.~\ref{fig:cat_results}b for 80 trajectories at each initial condition, where the radial direction indicates cat-state amplitude, $|\alpha|$, the polar coordinate gives cat state phase, $\phi$, and the color shows the number of node-teleportations from the the initial conditions (black dots denoted by arrows). (a) and (c) depict regular teleportation through a cluster state where cat states have initial amplitude $\alpha=2$ and $\alpha=4$, respectively. (b) and (d) have the same initial conditions as at left, but now PhANTM is performed as opposed to regular teleportation. A random (but known) phase kick is applied at each step by homodyne measurement in all cases, but one can see that PhANTM preserves cat-state amplitude up to a level dependent on cluster-state squeezing (here 15dB).}
    \label{fig:cat_phases}
    \end{figure}

Under this paradigm, including the PhANTM with a Gaussian cluster state can be seen as a machine that \textit{perpetuates} non-Gaussianity even in the presence of Gaussian noise. There exists a stable equilibrium at which some amount of non-Gaussianity will remain, which was previously only formulated for CV cluster states with fault-tolerant GKP error correction~\cite{Menicucci2014ft}. This protocol can thus work as a pseudo quantum memory for cat states when paired with a time-domain cluster state, as an input cat state can be teleported through many time-separated cluster-state nodes and retrieved later. The retrieved state will have randomized but known parameters dependent on specific measurement results at each step, and the retrieved state will remain within the cat-state manifold.\\ 
\indent This is demonstrated more clearly in Fig.~\ref{fig:cat_phases}, which depicts the same data as shown in Fig.~\ref{fig:cat_results}(b). Fig.~\ref{fig:cat_phases} shows the evolution of the phase, $\phi$, as the polar coordinate and amplitude, $|\alpha|$, as the radial value according to the fit with a cat state of the form of Eq.~\ref{eq:cat} for cat states as they are teleported across a cluster state. In each plot, the large black dot shows the parameters of the initial cat state introduced to the cluster which was taken to be an even cat ($\phi=0$) with no squeezing and amplitude of $\alpha=2$ (a and b) or $\alpha=4$ (c and d). In Figs.~\ref{fig:cat_phases}(a) and (c), the cat states were simply teleported normally using $P-$basis homodyne measurements. Figs.~\ref{fig:cat_phases}(b) and (d) begin with the same initial cat state but depict the effects of including PhANTM during teleportation across the cluster. As the states were teleported further along the cluster state, as indicated by color to represent node-distance, the amplitude of the cat states decayed as seen by the final magenta dots being clustered about the origin in Figs.~\ref{fig:cat_phases}(a) and (c). When the PhANTM was included, the cat-state amplitudes at no point spiraled to the center, but instead became clustered about an orbit at radius $|\alpha|\approx 1.5$, thus preserving the cat state and the associated non-Gaussianity. \\
\indent At each teleportation, finite-squeezing effects introduce Gaussian noise that dampens the amplitude of the cat state and introduces measurement-dependent randomization to the phase, which is also dependent on the amplitude of the cat state. This effect can be understood by realizing that as the cat-state amplitude increases, the interference fringes between the classical coherent state portions that determine the phase, $\phi$, oscillate at a higher frequency. Due to finite squeezing, the teleportation step applies a Gaussian envelope not centered about the origin in phase space, but instead centered about the measurement outcome. Shifts to this Gaussian envelope cause a slight `scanning' effect of the interference fringe, which changes the phase at each step more drastically for larger cat states. Because this phase-randomization is known from previous homodyne measurement results, a feed-forward displacement can be applied to shift the cat and realign the fringes to reset the phase. \\
\indent Examining Fig.~\ref{fig:cat_phases}(a) and (c), we can also see that the amplitude decays to zero through the poles of the plot when $\phi=0$ or $\phi=\pi$. This can be understood by realizing that as the cat shrinks, the interference fringes become dampened and very wide, so that a phase cannot be easily determined when the oscillations begin to vanish. These are the two extremes of shrinking cat states:  either the even ($\phi=0$) cat, which must tend toward the vacuum state as $|\alpha|<<1$, or the odd ($\phi=\pi$) cat, which tends toward a squeezed single photon. \\
\indent Additionally, it is important to note that finite squeezing dictates a threshold at which the cat-state amplitude saturates. If one wishes to achieve non-Gaussianity beyond this threshold in a given cluster, e.g., larger amplitude cats, it will be necessary to make use of distillation-type protocols. Many such protocols exist to take cat states and breed them to generate more non-Gaussianity~\cite{Lund2004,suzuki2006practical,laghaout2013amplification,etesse2014proposal,etesse2015experimental,Sychev2017,Vasconcelos2010,Weigand2018}, and we will show later how analogous protocols can be developed to breed cat states within the cluster-state architecture.

\subsubsection{Higher-dimensional Clusters}
\begin{figure}[!htb]
 \includegraphics[width = 0.45\textwidth]{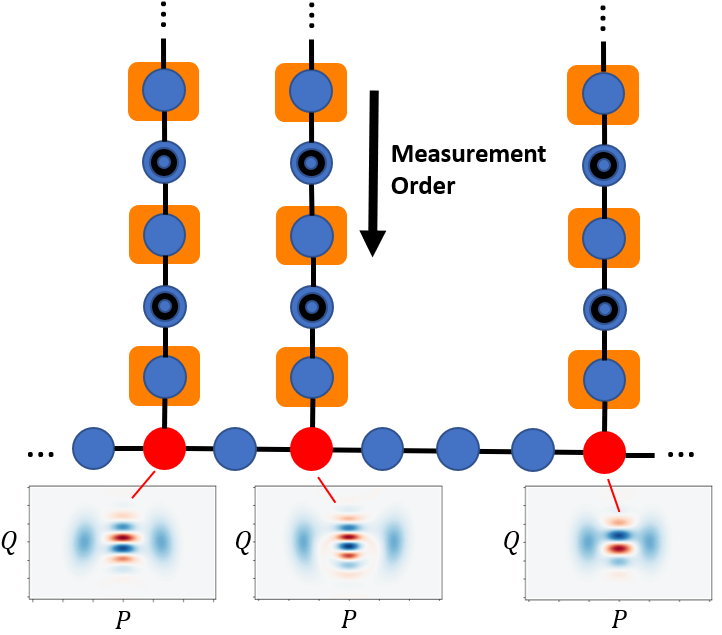}
  \caption{Measurements on 1-D quantum wires can be used to generate cat states embedded within a higher-dimensional cluster state. Orange blocks indicate a step of the PhANTM algorithm (attempted photon subtraction, homodyne detection, and feed-forward displacement) and black rings indicate $P$-basis homodyne measurements. The red nodes represent fully de-Gaussified cat states.}
    \label{fig:2dcluster}
    \end{figure}

As demonstrated above, the PhANTM can be used to transform a 1-D quantum wire of sufficient length into a cat state. A simple extension of this idea is to apply the algorithm on any 1-D topological wire connected to a higher dimensional cluster state to produce a cat state embedded within the remainder of the cluster. With this method, the algorithm can be performed simultaneously on many quantum wires to engineer a non-Gaussian cluster state with cat states embedded at strategic locations. Each of these cat states will have an average amplitude dependent on the initial cluster-state squeezing and the number of nodes in each 1-D wire consumed by the PhANTM algorithm. Gaussian operations can then be performed as per the usual cluster-state formalism with homodyne measurements to manipulate the exotic quantum states as desired, such as perhaps breeding the cat states as we will discuss further in the next section.
\\
\indent The process of embedding cat states in a higher-dimensional cluster can be visualized by Fig.~\ref{fig:2dcluster}, where a 2D cluster state is reduced to a 1-D cluster peppered with cat states. The PhANTM can be applied to all chains simultaneously, but measurements within each chain must proceed in the correct order --- top down in this example. \\
\indent It should be emphasized that without loss of generality, we can consider the local embedded state at each step in the cluster state as the state in the circuit before the application of $\opCZ$ gates. Because all $\opCZ$ gates commute, the order does not matter. This is one of the fundamental principles of measurement-based QC~\cite{Raussendorf2003}. Thus in Fig.~\ref{fig:2dcluster}, even though all $\opCZ$ gates are applied at the beginning, the localized structure of the entanglement in the graph means that the $\opCZ$ gates in the horizontal wire can be commuted to the end so that the $\opCZ$ gates and measurements in the vertical wires proceed and affect the local quantum states of the nodes embedded in the horizontal wire as if the horizontal $\opCZ$ gates had not yet been applied.\\
\section{Breeding Cat States}
\label{sec:breeding}

It has long been known that cat states can be probabilistically bred into larger amplitude cat states~\cite{Lund2004,suzuki2006practical,laghaout2013amplification,etesse2014proposal,etesse2015experimental,Sychev2017}, and recently it was demonstrated that a supply of squeezed cat states can be deterministically bred into a class of grid states~\cite{Weigand2018}. These procedures were formulated in terms of beamsplitter interactions, but each method can proceed analogously by performing homodyne measurements on cat states embedded in a cluster state. Instead of teleporting through a momentum eigenstate, each cat state is teleported through another cat state to produce the desired effect. Just as teleporting through a finite squeezed state leads to filtering of the teleported state with a Gaussian envelope~\cite{Menicucci2006,Alexander2014}, teleporting through an ancillary squeezed cat state filters the $P$-basis wavefunction of the input with the $Q$-basis wavefunction of ancilla shifted by the homodyne result. This is seen generically by replacing the momentum eigenstate from the teleportation circuit in Eq.~\ref{eq:canon_telep} with an arbitrary quantum state, $\ket{\psi'}$. This can be rewritten as an operator in $\op{Q}$ applied to a zero momentum eigenstate and commuted through the $\opCZ$ gate to have the general circuit
\begin{equation}\label{eq:gen_telep}
\begin{split}
\raisebox{-1.5em}{$\ketsub{\psi}{out}=$}
\hspace{10mm}
    \Qcircuit @C=0.5em @R=1em {
\lstick{\brasub{0}{P}}&\qw &\qw& \ctrl{1} &\gate{Z^\dag(m)} &\rstick{\ket{\psi}} \qw \\
\lstick{} &\qw&\gate{\sqrt{2\pi}\psi_Q'(Q)}&\control \qw &\qw&\rstick{\ketsub{0}{P}} \qw
%&\qw&\control \qw &\rstick{\ketsub{0}{P_2}} \qw
}\raisebox{-1.5em}{\hspace{8mm}.}
\end{split}
\end{equation}
Pulling the circuit taut and commuting the resultant Fourier transform to the end, we have that
\begin{equation}
    \ketsub{\psi}{out}=\sqrt{2\pi}\op{R}(\tfrac{\pi}{2})\psi'_Q(-\op{P})\op{Z}^\dag(m)\ket{\psi},
\end{equation}
where the subscript on $\psi'_Q(-\op{P})$ indicates that it is the $Q$-basis wavefunction of the starting ancillary state, even though it is a function of the $\op{P}$ operator. Finally, commuting the displacement through to the left reveals the result of
\begin{equation}
    \ketsub{\psi}{out}=\sqrt{2\pi}\op{X}(m)\op{R}(\tfrac{\pi}{2})\psi'_Q(-\op{P}+m)\ket{\psi}.
    \label{eq:any_input_telep}
\end{equation}
This shows that up to an overall displacement and rotation, the initial $P$-basis wavefunction for $\ket{\psi}$ transforms as $\psi_P(x)\rightarrow \psi_P(x)\psi'_Q(m-x)$. Thus, in the $P$-basis, $\ket{\psi}$ is filtered by the $Q$-wavefunction of $\ket{\psi'}$.
\\
\indent For the sake of illustration, supposed we input two equivalent squeezed cat states into the circuit but with one Fourier transformed with respect to the other, so that their wavefunctions are each proportional to the sum of two Gaussian peaks,
\begin{equation}
\scalebox{0.96}{$
    \psi_P(x)=\psi'_Q(x)= \mathcal{N}
    \left(e^{-\frac{1}{2s}(x-\alpha)^2}+e^{-\frac{1}{2s}(x+\alpha)^2}\right)$},
\end{equation}
where
\begin{equation}
    \mathcal{N}=\frac{\pi^{-1/4}\sqrt{s}}{\sqrt{2+2e^{-2|\alpha|^2}}}.
\end{equation}
Supposing the measurement result is $m=0$, we have that up to an overall rotation, the new wavefunction, $\phi$, is
\begin{align}
    \phi_P(x)&=\psi_P(x)\psi'_Q(x)\nonumber\\
    &\propto e^{-\frac{1}{s}(x-\alpha)^2}+e^{-\frac{1}{s}(x+\alpha)^2}+2e^{-\frac{1}{s}(x^2+\alpha^2)}.
\end{align}
 For peak separations of $\alpha\gtrsim 1$, the third term becomes negligible, and the wavefunction is the same as the initial wavefunction $\psi$, but with narrower peaks. If we were to apply a squeezing operation to bring the peak widths back to the starting width, the peak separation would increase, and it is evident that this is just a larger cat state than what we started with. The above process is an example of breeding to enlarge cat states.
 
\begin{figure}[H]
\centering
 \includegraphics[width = 0.35\textwidth]{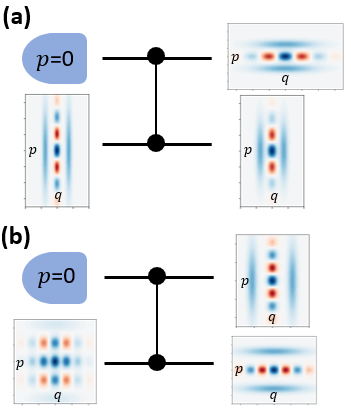}
 \caption{Cat states can be bred by teleporting one through the other to produce larger amplitude cat states as shown in (a) or grid states as shown in (b). The type of bred state is dependent on input cat state orientation.}
    \label{fig:cat_breed}
\end{figure}

 Now, suppose the inputs were the same cat states as above, but first each mode was Fourier transformed so that
 \begin{equation}
\scalebox{0.96}{$
    \psi_P(x)=\psi'_Q(x)= \mathcal{N}
    e^{-\frac{1}{2}s^2x^2}\left(2\cos{\alpha x}\right)$}.
\end{equation}
Again taking a measurement outcome of $m=0$ for illustration,
the output wavefunction, $\phi$, from this process before the rotation would be
\begin{equation}
  \phi_P(x)\propto e^{-s^2x^2}\left(\cos{\alpha x}\right)^2,  
\end{equation}
which in the $Q$-basis is
\begin{equation}
    \phi_Q(x)\propto \left(e^{-\frac{1}{s}(x-\alpha)^2}+e^{-\frac{1}{s}(x+\alpha)^2}+2e^{-\frac{1}{s}x^2}\right).
\end{equation}
The above equation shows that with this case, the final state is now a superposition of three peaks with a binomial distribution as opposed to the two-peak superposition of the starting cat states. One can see how repeating this process can eventually give rise to a state with many peaks in the superposition which can approximate a grid state. These two cases are depicted in Fig.~\ref{fig:cat_breed}, where the process at left breeds to enlarge cats and the process at right breeds to create grids.
As one can see, the output state depends on the phase of the input cat states, which is known from the iterated preparation method and can be controlled by performing regular teleportations to enact $\op{R}(\pi/2)$ gates. 
Cat-state breeding has been described in detail elsewhere~\cite{Lund2004,suzuki2006practical,laghaout2013amplification,etesse2014proposal,etesse2015experimental,Sychev2017, Vasconcelos2010, Weigand2018}, but the point of interest here is that it can be done \textit{within} the cluster state using only Gaussian measurements once all initial cats have be embedded within the cluster. Previous methods have examined breeding with beamsplitters and homodyne measurements in great detail, and we show in Appx.~\ref{Appx:BS2CZ} how the beamsplitter breeding maps directly to breeding with $\opCZ$ gates, where the only difference is that one input must be first Fourier transformed and the output has an additional Gaussian operation that can be undone with feed-forward displacement and Gaussian information processing on the cluster state. This mapping implies that all previous results based on breeding cat states with beamsplitters holds within the canonical cluster state.

One may ask why including a probabilistic scheme to enlarge cats is necessary when we have already demonstrated a scheme to generate cat states with near-unity success after enough iterations. However, it is important to remember that the mean amplitude of cat state produced by the PhANTM protocol is limited in part by the squeezing present in the cluster state. Breeding, however, allows for one to have a chance at sacrificing pairs of smaller cat states to enlarge them. By performing PhANTM in paralell on several 1-D chains, one could foreseeable have enough weak cats that it would be advantageous to attempt breeding them into larger resource states. 

\subsection{GKP States}
Instead of taking a pair of cat states and breeding to generate a larger cat state, one can breed with a different outcome in mind, such as the generation of grid states. For this purpose, we more closely examine the case at right in Fig.~\ref{fig:cat_breed}, which when repeated, can succeed without post-selection as described previously~\cite{Weigand2018}.% {\color{red} [RNA: transition from breeding cats to breeding GKP states was a bit confusing]} 
We provide the Kraus operator and resultant state after an arbitrary breed step in Appx~\ref{Appx:breed}.

The process of performing alternating PhANTM steps as detailed in the previous section acts to produce cat states on the cluster state, but breeding useful grid states will require squeezed cats. This is no obstacle, however, as any single-mode Gaussian operation can be implemented with a series of four homodyne measurements on the cluster~\cite{Ukai2009}. For the specific case of squeezing, it suffices to apply three successive shear gates of $e^{\frac{i\gamma_k}{2}\op{Q}^2}$ with properly selected $\gamma_k$ to effectively apply the gate $\op{S}(\sq)$. For squeezing $s=e^\sq$, the values of $\gamma_k$ are $\gamma_1=\gamma_3=s$ and $\gamma_2=s^{-1}$  Note that the maximum squeezing operation that can be applied in this way is limited by the squeezing level of the cluster.

\begin{figure}[h]
 \includegraphics[width = 0.45\textwidth]{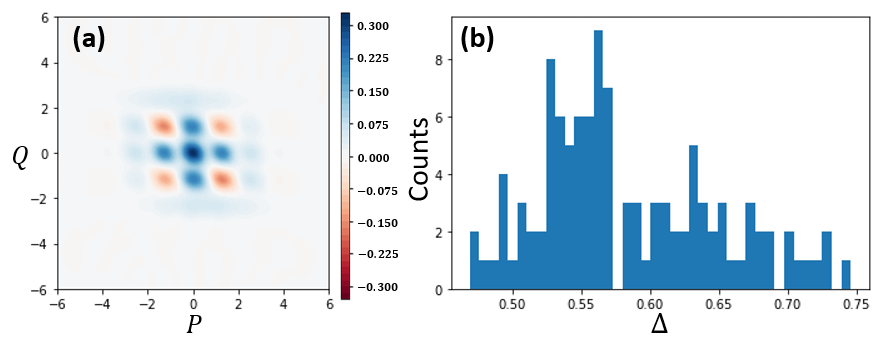}
  \caption{(a) Average GKP qunaught obtained by performing Gaussian operations and measurements within the cluster state on a pair of states generated from the PhANTM algorithm. (b) Histogram of $\Delta$ fit parameters for the GKP qunaughts bred from all states obtained by the PhANTM algorithm.}
    \label{fig:grid_data}
\end{figure}

Suppose we begin with a cluster state consisting of several quantum wires and perform many steps of the PhANTM protocol on each quantum wire until we have successfully generated a cat state, which is now entangled to the remaining portion of the cluster state. Each of these can be squeezed through a series of three homodyne measurements, and then pairs can be bred to make grid states without post-selection. An example of the resultant grid state is shown in Fig.~\ref{fig:grid_data}(a), where we have used states generated from average results after $M=35$ steps of the PhANTM protocol in a cluster of $17$ dB squeezing, squeezed each one with three shear operations, and bred the result. Gaussian noise due to finite squeezing is included in all parts of the calculation to simulate realistic application of information processing on the cluster state. 

To have a benchmark with some consistency, the target state was chosen to be the GKP qunaught, which is the GKP state with equal grid spacing in both quadratures. In order to make this state, the spacing of the initial cats must be correctly chosen so that the final spacing of the grid will be symmetric. This was achieved by tuning the shearing parameters used for the applied squeeze operation. 

The asymmetries in the Wigner function shown arise from the stochastic homodyne measurements used when applying the squeezing with cluster-state processing. The final homodyne measurement for the breed step was post-selected on zero to achieve consistent states within the approximate GKP family to be used for direct comparison of the quality of the state. Varying the measurement result will not change final grid spacing or width in the $Q$-quadrature, but the introduced phase will alter the symmetry between the quadratures. Nevertheless, a grid state will always be created under asymptotic breeding regardless of measurement results~\cite{Weigand2018}. We repeated this procedure for each cat state created with the PhANTM algorithm using the same data as shown in Fig.~\ref{fig:cat_results}(a). Each cat state was squeezed with quantum information processing on the cluster state, and then bred with a copy to produce a similar grid state to the one shown. The resultant state was then fitted to a GKP qunaught with varying peak widths, where $\Delta^2$ is the variance of each peak with the squeezing of the GKP state in dB given by $s_{GKP}=-10\log_{10}\left(2\Delta^2\right)$. The resulting histogram is shown in Fig.~\ref{fig:grid_data}(b), where the fidelity with the matched grid state averaged above $0.95$, but was above $0.98$ for all states with $\Delta<0.6$. The values of $\Delta$ shown in the histogram are well above what is needed for fault-tolerance~\cite{Menicucci2014ft,Fukui2018}, but this is unsurprising for a single breeding round. It has been shown that efficient cat-state breeding schemes can lead to GKP states that are more strongly squeezed than the cat states used to create them~\cite{Weigand2018}.

With sufficiently large cat states and high squeezing in the cluster state, breeding within the cluster allows for the generation of a supply of GKP states that are embedded within the computational resource. These can then be teleported throughout the cluster where they can then be used for fault-tolerant error correction~\cite{Menicucci2014ft,Walshe2019} and universal QC~\cite{Baragiola2019}, or alternatively, states can be teleported through the embedded GKP state to enact error correction directly~\cite{knill2005scalable}. Thus, while GKP states may be essential for error correction on cluster states, cluster states may also be valuable ingredients for the synthesis of GKP states.

\subsection{Interpretation as Phase Estimation}
\label{sec:phast_est}
In this section, we wish to deviate briefly from the main results to make a pedagogical connection to other techniques used in bosonic encodings such as phase estimation. Quantum phase estimation for unitary operator $\op{U}$ is a procedure to measure the eigenvalue of the operator, $e^{i\theta}$, and project the input state into the corresponding eigenstate, $\op{U}\ket{\psi_{\theta}}=e^{i\theta}\ket{\psi_{\theta}}$. Phase estimation, which can proceed through several methods~\cite{svore2018faster}, was proposed to be implemented with ancilla qubits to generate GKP states~\cite{terhal2016encoding}. This has been successfully demonstrated with microwave cavity fields and transmon qubits~\cite{CampagneIbarcq2020}. A similar phase-estimation scheme without qubits was developed to create GKP states from cat states~\cite{Weigand2018}, which is the method we connect to above for breeding cat states. We now complete the connection to phase-estimation on a cluster state for unitary operators that are phase-space displacements. 

For a general unitary operation $\op{U}=e^{i\op{\mathcal{T}}}$ where $\op{\mathcal{T}}$ is a time-independent Hermitian operator (which we will later take to be proportional to $\op{P}$), a single round of phase estimation can be implemented using an ancillary qubit described by the circuit
\begin{equation}\label{eq:phase_est_qubit}
\begin{split}
%\hspace{20mm}
    \Qcircuit @C=0.5em @R=1em {
\lstick{(\text{out})}&\qw & \gate{Ue^{i\varphi}} &\qw &\rstick{(\text{in})} \qw \\
 \lstick{\brasub{\pm}{qubit}}&\qw&\ctrl{-1} \qw &\qw&\rstick{\ketsub{+}{qubit}} \qw
%&\qw&\control \qw &\rstick{\ketsub{0}{P_2}} \qw
}\raisebox{-1em}{\hspace{8mm},}
\end{split}
\end{equation}
where $e^{i\phi}$ is a phase factor. Here, a controlled-$\op{U}e^{i\varphi}$ is applied to the input state state in mode one where a qubit initialized to the $\ket{+}$ state is used as a control. This controlled unitary can be applied by implementing the operator
\begin{align}
    &C_{U_{12}}=e^{\frac{i}{2}(\op{\mathcal{T}}_1+\varphi)}e^{-\frac{i}{2}\left((\op{\mathcal{T}}_1+\varphi\id_1)\otimes\op{\sigma}_{z_2}\right)}\nonumber \\
    &= e^{\frac{i}{2}(\op{\mathcal{T}}_1+\varphi)}\left(\cos(\tfrac{\op{\mathcal{T}}_1+\varphi}{2})\otimes \id_2-i\sin(\tfrac{\op{\mathcal{T}}_1+\varphi}{2})\otimes\op{\sigma}_{z_2})\right).
    \end{align}
The ancilla qubit is then measured in the $\pm$-basis. Consider the case when the applied unitary is a displacement. Without losing the generality of arbitrary displacements, we can take
\begin{equation}
    \op{U}=e^{i\alpha\op{P}},
\end{equation}
and note that rotating the input state before and after applying $\op{U}$ will give the desired freedom for arbitrary displacements. In this case, the effective measurement-based operator applied to the input state to the circuit in Eq.~\ref{eq:phase_est_qubit} is given by
\begin{align}
    \op{\mathcal{M}}_+&=e^{\frac{i}{2}(\alpha\op{P}_1+\varphi)}\cos(\tfrac{\alpha \op{P}+\varphi}{2})\\
    \op{\mathcal{M}}_-&=ie^{\frac{i}{2}(\alpha\op{P}_1+\varphi)}\sin(\tfrac{\alpha \op{P}+\varphi}{2}).
\end{align}
Thus up to a global phase and a displacement, a single round of phase estimation applies a sinusoid in $\op{P}$ with an additional measurement-dependent phase shift of either $0$ or $\tfrac{\pi}{2}$.

We can now compare the result of this circuit to teleporting an input through a squeezed cat state on a cluster state using the circuit
\begin{equation}\label{eq:phase_est_cluster}
\begin{split}
\hspace{20mm}
    \Qcircuit @C=1em @R=1.5em {
\lstick{\brasub{m}{P}}&\qw & \ctrl{1}&\qw  &\rstick{(\text{in})} \qw \\
\lstick{(\text{out})} &\qw&\control \qw &\qw&\rstick{\ket{cat}} \qw
%&\qw&\control \qw &\rstick{\ketsub{0}{P_2}} \qw
}\raisebox{-1em}{\hspace{12mm}.}
\end{split}
\end{equation}
If the cat state is oriented such that the fringes are along the $P$-quadrature, that is if 
\begin{equation}
    \ket{cat}\propto \int\!dxe^{-x^2/2s^2}\cos(\alpha x)\qket{x}
\end{equation}
where $s=e^\sq$ is the squeezing,
then using the results of the general teleportation circuit given by Eq.~\ref{eq:any_input_telep}, shows that this circuit applies the operator
\begin{equation}
    \op{\mathcal{M}}_m=\sqrt{2\pi}\op{X}(m)\op{R}(\tfrac{\pi}{2})e^{-\op{P}^2/2s^2}\cos(\alpha\op{P}-m).
\end{equation}
After undoing the measurement-induced displacement and rotation with further cluster-state processing, and for large enough squeezing, this operator reduces to
\begin{equation}
    \op{\mathcal{M}}_m\approx\sqrt{2\pi}\cos(\alpha\op{P}-m),
\end{equation}
which is identical to the qubit-based phase estimation up to an overall displacement, where the applied sinusoid operation now has a phase shift that is continuous based on the measurement result. By performing Fourier transforms before and after this circuit (through teleportation on the cluster state), the operator can instead be transformed to a sinusoid in $\op{Q}$.

Inserting another cat state for the input state in Eq.~\ref{eq:phase_est_cluster} is exactly the method of GKP systhesis as described in the previous section and completes the connection to Ref.~\cite{Weigand2018} where repeated phase-estimation is used to make grid states. More generally, we can borrow ideas from other bosonic phase-estimation schemes and enact them using cluster states and PhANTM generated cat states.

\section{Macronode Extension}
\label{sec:macronode}
Many experimental implementations of cluster states have not generated the canonical cluster based on $\CZ$ entangling gates, which would require inline squeezing, but instead rely on linear optics with all squeezing generated up-front~\cite{Yokoyama2013,Chen2014,Asavanant2019,Larsen2019}. These types of cluster states are formed from \textit{macronodes}, which are a collection of two or more physical modes containing non-locally distributed information encoding a single logical state~\cite{Alexander2014}. In this section, we develop the connection from the canonical case considered above to the macronode implementation that has been experimentally realized. We first briefly introduce formalism and techniques for manipulating quantum information in the macronode cluster and then develop a dictionary protocol to show that PhANTM can be used to embed cat states into macronode cluster states as well.\\
\indent The macronode cluster states are created by beginning with many entangled resources in the form  of two-mode squeezed states, which can be generated either directly or by interfering two single-mode squeezed states on a beamsplitter, and then entangled pairs are linked up through more beamsplitter interactions. Quantum information processing can proceed analogously to the canonical case, where now all physical modes within each macronode are subject to homodyne measurements. 

The simplest macronode cluster state, the quantum wire, contains two physical modes per macronode and has been used to experimentally implement Gaussian operations~\cite{asavanant2020one}. Weaving together quantum wires can produce higher-dimensional cluster states~\cite{Wang2014a,alexander2018universal}; the 2D case of this has recently been used to implement a set of universal Gaussian gates~\cite{larsen2021deterministic}. The PhANTM algorithms discussed previously in the context of canonical cluster states can be translated to macronode clusters. \\
\indent Before giving the result, we first review the formalism recently developed to treat general teleportation on a macronode wire, where arbitrary and potentially non-Gaussian ancillary inputs can be used to impart non-Gaussianity on the cluster through teleportation~\cite{walshe2020continuous}. This operation can be described by a local teleportation `gadget' which has the circuit form: 

\begin{align}\label{eq:macro_gen_telep}
    \begin{split}
    \Qcircuit @C=1.75em @R=2em 
    {
	&\lstick{\brasub{m_1}{P_{\theta_1}}}   & \bsbal{1} & \qw & \rstick{\text{(in)}} \qw[-1] &  \\
	&\lstick{\brasub{m_2}{P_{\theta_2}}}  & \qw   &  \bsbal{1} & \rstick{\ket{\psi}} \qw \\
	&\lstick{\text{(out)}}	& \qw  &   \qw  & \rstick{ \ket{\phi} } \qw \\
		}\, 
	\end{split}	
\end{align}
where in this formalism, input states travel from right to left. We also adopt the convention that an arrow represents a balanced beamsplitter interaction between two qumodes,

\begin{equation}\label{BScircuit}
\begin{split}
    \Qcircuit @C=1em @R=2.1em {
	   & \bsbal{1} & \rstick{j} \qw  &&&&&&&&&&\\
	 &\lstick{} \qw       & \rstick{k} \qw&&&&&& &\raisebox{2.5em}{= $\bsop_{jk}(\tfrac \pi4) :=e^{\frac{\pi}{4}(a_j a_k^\dag - a_j^\dag a_k)}$}. \\
		}
\end{split}		
\end{equation}
The overall effect of this teleportation circuit is given by 

\begin{equation}\label{taut_circ}
\begin{split}
\resizebox{.71\columnwidth}{!}{
    \Qcircuit @C=0.75em @R=1em {
        &\lstick{\text{(out)}} &\gate{\frac{1}{\pi}A({\psi,\phi}){}} &\gate{ D(\mu) } &\gate{V(\theta_1 ,\theta_2)} &\rstick{\text{(in)}}\qw \\
	} 
	}
\end{split}
\end{equation}

which is the same as applying the Kraus operator,
\begin{equation}
    \op{K}_{m_1,m_2}= \frac{1}{\pi}\op{A}(\psi,\phi)\op{D}(\mu)\op{V}(\theta_1,\theta_2),
\end{equation}
to the input state. Here, $\op{D}(\mu)$ is a displacement based on the homodyne measurement with
\begin{equation}
    \mu=\frac{m_1 e^{i\theta_2}+m_2 e^{i\theta_1}}{\sin{(\theta_2-\theta_1)}}.
    \label{eq:mu}
\end{equation}
and $\op{V}$ is a Gaussian operation that depends only on the measurement basis given by %{\color{red} [RNA: I think you mean $\log \tan$ instead of $e^{\tan}$]}
\begin{equation}
    \op{V}(\theta_1,\theta_2)=\op{R}(\theta_+-\tfrac\pi2)\op{S}\left(\ln{(\tan{\theta_-})}\right)\op{R}(\theta_+).
    \label{eq:V_op}
\end{equation}

We have additionally defined
\begin{equation}
    \theta_{\pm}=\frac12(\theta_1 \pm \theta_2).
\end{equation}
The potentially non-Gaussian operation, $\op{A}(\psi,\phi)$, comes from non-Gaussianity within the input states $\ket{\psi}$ and $\ket{\phi}$, and can be determined by
\begin{equation}
    \op{A}(\psi,\phi)=\iint\! d^2\alpha\widetilde{\psi}(\alpha_I)\phi(\alpha_R)\op{D}(\alpha)
\end{equation}
with complex parameter $\alpha=\alpha_R+i\alpha_I$. The respective position and momentum wavefunctions of the ancillary states are given by $\widetilde{\psi}(t)={}_P\!\langle t \ket{\psi}$ and $\phi(s)={}_Q\!\langle s \ket{\phi}$. In the case where the input states are Gaussian only, then $\op{A}(\psi,\phi)$ is also Gaussian. In the specific case where the inputs are position and momentum eigenstates, $\ket{\psi}=\ket{0}_P$ and $\ket{\phi}=\ket{0}_Q$, then the operator $\op{A}(\psi,\phi)$ reduces to the identity, and we have the ability to apply a complete Gaussian gate set as in the canonical case.

Derived in Ref.~\cite{walshe2020continuous}, we will make use of the identity

\begin{equation} \label{eq:gaus_proj}
\begin{split}
\centering
\resizebox{.8\columnwidth}{!}{
 \Qcircuit @C=1em @R=1.5em {                                              
           &\qw&\rstick{\ketsub{m_1}{P_{\theta_1}}} \qw &&&&&&\qw&\gate{\frac{1}{\sqrt{\pi}}V^{\dag}(\theta_1, \theta_2)} 
           &\gate{D^{\dag}(\mu)} &\ar @{-} [dr(0.5)] \qw  &\\
          &\bsbal{-1}  \qw& \rstick{\ketsub{m_2}{P_{\theta_2}}} \qw &&&&\raisebox{2.5em}{$=$} && \qw &\qw                                   &\qw                  &\ar @{-}[ur(0.5)]
          \qw  & 
    } \; \raisebox{-1em}{,}
    }
\end{split}
\end{equation}

where the ended circuit represents an EPR pair,

\begin{equation}  \label{eq:EPR_def}
       \begin{split}
 \Qcircuit @C=0.2cm @R=0.3cm {
         &&&&&&&&&  \qw & \ar @{-} [dr(0.5)]\qw  &\\
         &&\ustick{\raisebox{.1em}{$\smash{\ket{\text{EPR}}} \;\coloneqq$}}&&&&&&& \gate{\frac{1}{\sqrt{2\pi}} I } & \ar @{-} [ur(0.5)] \qw  & 
    } 
       \end{split}\, ,
\end{equation}
and we define the EPR pair in the q-basis as
\begin{equation}
    \label{eq:epr}
    \ket{\text{EPR}}_{12}=\frac{1}{\sqrt{2\pi}}\int\!dt\ket{t}_{1Q}\ket{t}_{2Q}.
\end{equation}

One particular use of a diagrammatic approach is that operators can be easily transferred, or `bounced', from one mode to the other across an EPR pair~\cite{walshe2020continuous}. This bouncing can be represented as

\begin{align}\label{eq:bouncing}
    \begin{split}
    \Qcircuit @C=1em @R=1em {
    &\qw & \gate{O} & \qw[-1] \ar @{-} [dr(0.5)] & & & &\qw & \qw & \qw[-1] \ar @{-} [dr(0.5)] &\\
    &\qw & \qw & \qw[-1] \ar @{-} [ur(0.5)] & & \raisebox{2em}{$=$} & &\qw & \gate{O^{\text{T}}} & \qw[-1] \ar @{-} [ur(0.5)] & \, \\
    }
    \end{split},
\end{align}

where the operator $\hat{O}$ acting on one mode of the EPR pair is the same as $\hat{O}^T$ acting on the other, where the transpose of the operator is taken in the Q-basis, so that $\op{Q}^T=\op{Q}$ while $\op{P}^T=-\op{P}$. This formalism is useful whenever two homodyne measurements are performed on neighboring entangled nodes.

The method we have described to generate cat states on a Gaussian cluster is predicated on finding a means to apply polynomial operators comprised solely of the $\op{Q}$ operator (or solely of $\op{P}$, up to a Fourier transform) to the initial squeezed vacuum nodes of the cluster. In the canonical case, this was achieve by adding a photon subtraction step directly before the measurement. We now develop here the associated analogy with macronodes by making use of the dictionary protocol~\cite{Alexander2014}.

\subsection{Dictionary Protocol}
Consider a macronode teleporation circuit where we have applied some operator $\op{O}$ before the first homodyne measurement, and the measurement angles are chosen such that we perform respective $P$ and $Q-$basis measurements ($\theta_1=0, \theta_2=\tfrac{\pi}{2}$). With the inclusion of finite squeezing, this circuit will appear as
\begin{align}\label{eq:macro_dict}
    \begin{split}
    \Qcircuit @C=1.5em @R=1.5em 
    {
	&\lstick{\pbra{m_1}}   & \gate{O}& \bsbal{1} & \qw &\qw& \rstick{\text{(in)}} \qw[-1] &  \\
	&\lstick{\qbra{m_2}}  & \qw   & \qw &\ctrl{1} &\gate{S(\sq)}& \rstick{\ketsub{0}{N}} \qw \\
	&\lstick{\text{(out)}}	& \gate{R(\tfrac{\pi}{2})} & \qw  \raisebox{3em}{$\scriptstyle{\tanh{2\sq_0}}$}&\control  \qw &\gate{S(\sq)}&  \rstick{ \ketsub{0}{N} } \qw \\
		}\, 
	\end{split} \hspace{4mm}.	
\end{align}
Note that for the formation of the finite-energy EPR pair in the bottom two modes we have used a weighted $\opCZ$ gate of $e^{i\tanh{2\sq_0}\op{Q}_2\op{Q}_3}$, which is the same as the finite-squeezing version of the beamsplitter case in Eq.~\ref{eq:macro_gen_telep}, up to a rotation on the output and the second input squeezed state, and a re-scaling of the initial squeezing. If the single-mode squeezed states before the entangling beamsplitter have squeezing $\sq_0$, then the effective squeezing before the weighted $\opCZ$ gate will be $\sq=\ln\sqrt{\text{sech}{2\sq_0}}$. The weighting on the $\opCZ$ gate can be returned to one by noting that
\begin{equation}
\resizebox{.9\columnwidth}{!}{
    $e^{i\tanh{2\sq_0}\op{Q}_2\op{Q}_3}=\op{S}^\dag_3(\ln[\tanh{2\sq_0}]) e^{i\op{Q}_2\op{Q}_3}\op{S}_3(\ln[\tanh{2\sq_0}])$}.
\end{equation}
With this, the circuit becomes
    \begin{align}
    \begin{split}
    \centering
    \resizebox{.9\columnwidth}{!}{
    \Qcircuit @C=1.5em @R=1.5em 
    {
	&\lstick{\pbra{m_1}}   & \gate{O}& \bsbal{1} & \qw &\qw& \rstick{\text{(in)}} \qw[-1] &  \\
	&\lstick{\qbra{m_2}}  & \qw   & \qw &\ctrl{1} &\gate{S(\sq)}& \rstick{\ketsub{0}{N}} \qw \\
	&\lstick{\text{(out)}}	& \gate{R(\tfrac{\pi}{2})} & \gate{S^\dag\left(\scriptstyle{\ln(\tanh{2\sq_0})}\right)} \qw &\control  \qw &\gate{S(\sq')}&  \rstick{ \ketsub{0}{N} } \qw \\
		}}
	\end{split} \hspace{4mm},
\end{align}
where $\sq'=\ln\left(\tanh{2\sq_0}\sqrt{\text{sech}{2\sq_0}}\right)$. We now derive the specific dictionary protocol that will determine the effect of applying $\op{O}$ as if it were applied to a canonical cluster.
We start by commuting the beamsplitter with the $\opCZ$ gate using the identity
\begin{equation}
    \op{B}_{12}(\tfrac{\pi}{4})\op{C}_{Z_{23}}\op{B}_{12}^\dag(\tfrac{\pi}{4})=e^{\tfrac{i}{\sqrt{2}}(\op{Q}_2-\op{Q}_1)\op{Q}_3},
\end{equation}
which is just two weighted $\opCZ$ gates. We can act this to the left on the measurement on mode two and fix the weight of the remaining $\opCZ$ by rotating the operator in the Heisenberg picture with squeezing operators. This is written as
\begin{align}
    &\brasub{m_2}{2Q}e^{\tfrac{i}{\sqrt{2}}(\op{Q}_2-\op{Q}_1)\op{Q}_3}=\brasub{m_2}{2Q}\op{Z}_3(\tfrac{m_2}{\sqrt{2}})e^{-\tfrac{i}{\sqrt{2}}\op{Q}_1\op{Q}_3} \nonumber \\
    =&\brasub{m_2}{2Q}\op{Z}_3(\tfrac{m_2}{\sqrt{2}})\op{R}_3(\pi)\op{S}_3(\tfrac{1}{2}\ln{2})\op{C}_{Z_{13}}\op{S}_3^\dag(\tfrac{1}{2}\ln{2})\op{R}^\dag_3(\pi).
\end{align}

The rotations can both be pushed to the outer edges of the circuit, where a rotation of $\op{R}^\dag_3(\pi)$ to the right has no effect, and the rotation of $\op{R}_3(\pi)$ on the left changes the sign of the $P$-quadrature displacement and combines with the end rotation. This results in the circuit
\begin{align}
\label{eq:macro_reduced_circuit}
    \begin{split}
    \Qcircuit @C=1em @R=1em 
    {
	&\lstick{\pbra{m_1}}   & \gate{O}& \ctrl{2}&\qw&\bsbal{1} &\qw& \qw&\rstick{\text{(in)}} \qw[-1] &  \\
	&&&&\lstick{\qbra{m_2}}  & \qw &\qw&\gate{S(\sq)}& \rstick{\ketsub{0}{N}} \qw \\
	&\lstick{\text{(out)}}	& \gate{G} &\control \qw &\gate{S(\sq'-\tfrac{1}{2}\ln{2})}& \qw& \qw& \qw& \rstick{ \ketsub{0}{N} } \qw \\
		}\, 
	\end{split}\hspace{6mm}	
\end{align}
where the Gaussian operation $\op{G}$ is
\begin{equation}
    \op{G}=\op{R}^\dag(\tfrac{\pi}{2})\op{Z}^\dag(\tfrac{m_2}{\sqrt{2}}{\scriptstyle\tanh{2\sq_0}})\op{S}^\dag\left(\scriptstyle{\ln(\tfrac{1}{\sqrt{2}}\tanh{2\sq_0})}\right).
\end{equation}

%{\color{red} [RNA: Can we relate this to equation 23 in Ref.~\cite{alexander2018universal}?]}
From the above circuit, it is easy to see that if we can reduce the beamsplitter and measurement on the second wire to a single Kraus operator acting on the top wire, then we have a macronode teleportation circuit with some additional operator acting on the input and a Gaussian operator $\op{G}$ on the output. To do this, we need to find the operator for
\begin{equation}
\label{eq:m2c_bs_op}
    \brasub{m_2}{2Q}\op{B}_{12}(\tfrac{\pi}{4})\op{S}_2(\sq)\ketsub{0}{2N}.
\end{equation}
Using Eq.~\ref{eq:sq_trick}, the fact that $\op{B}_{12}(\theta)=\op{B}_{12}^\dag(-\theta)=\op{B}_{21}(-\theta)$, and the beamsplitter decomposition given in Ref.~\cite{walshe2020continuous} of
\begin{equation}
    \op{B}_{j,k}(\tfrac{\pi}{4})=e^{i\op{P}_j\op{Q}_k}\op{S}_j(\tfrac{1}{2}\ln{2})\op{S}^\dag_k(\tfrac{1}{2}\ln{2})e^{-i\op{Q}_j\op{P}_k},
\end{equation}
we can rewrite Eq.~\ref{eq:m2c_bs_op} as
\begin{align}
\label{eq:m2c_bs_op2}
\pi^{1/4}\sqrt{\tfrac{2}{s}}&\op{X}^\dag_1(m_2)\op{S}_1(\tfrac{1}{2}\ln{2})\times \nonumber\\
&\brasub{m_2}{2Q}\op{S}_2^\dag(\tfrac{1}{2}\ln{2})e^{-i\op{Q}_1\op{P}_2}e^{-\op{Q}^2_2/2s^2}\ketsub{0}{2P},
\end{align}
where again $s=e^{\sq}$. The operators between the bra and ket in Eq.~\ref{eq:m2c_bs_op2} can be commuted to have all operators in $\op{Q}_2$ to the left and all operators in $\op{P}_2$ to the right, which gives
\begin{equation}
    e^{-\op{Q}^2_2/s^2} e^{\sqrt{2}\op{Q}_1\op{Q}_2/s^2}\op{S}_2^\dag(\tfrac{1}{2}\ln{2})e^{-\op{Q}^2_1/s^2}e^{-i\op{Q}_1\op{P}_2}.
\end{equation}
When acted on by the $Q$-basis bra and $P$-basis ket, along with noting that squeezing a quadrature eigenstate of zero has no effect, this reduces to an operator acting on wire one only; Eq.~\ref{eq:m2c_bs_op} becomes
\begin{equation}
\label{eq:m2c_bs_op3}
    \pi^{1/4}\sqrt{\tfrac{2}{s}}\op{X}^\dag_1(m_2)\op{S}_1(\tfrac{1}{2}\ln{2})e^{-\frac{(\op{Q}_1-\sqrt{2}m_2)^2}{2s^2}}.
\end{equation}
This expression is the same result as Eq.(23) in Ref.~\cite{alexander2018universal}, where the wavefunction of the input state used here from mode two is that of a squeezed vacuum state.

Taking Eq.~\ref{eq:m2c_bs_op3} and using Eq.~\ref{eq:sq_trick} on the third wire in the circuit to commute the effect of finite squeezing with the $\opCZ$ gate, we have that our circuit becomes 
\begin{equation}
    \Qcircuit @C=0.5em @R=1em {
&&&\lstick{\brasub{m_1}{P}} &\gate{O}&\ctrl{1} & \gate{X^\dag(m_2)} &\gate{S(\scriptstyle{\frac{1}{2}\ln{2}})} &\gate{e^{-\frac{(\op{Q}-\sqrt{2}m_2)^2}{2s^2}}}& \rstick{(\text{in})} \qw \\
&&&\lstick{\text{(out)}}&\gate{G}&\control \qw &\gate{e^{-Q^2/s'^2}} &\qw&\rstick{\pket{0}} \qw
}\raisebox{-1.5em}{\hspace{10mm},}
\end{equation}
with $s'=e^{\sq'}$.
For high initial squeezing, this circuit reduces to 
\begin{equation}
    \Qcircuit @C=0.5em @R=1em {
&&&\lstick{\brasub{m_1}{P}} &\gate{O}&\ctrl{1} & \gate{X^\dag(m_2)} &\gate{S(\scriptstyle{\frac{1}{2}\ln{2}})} & \rstick{(\text{in})} \qw \\
&&&\lstick{\text{(out)}}&\gate{G}&\control \qw &\qw &\qw&\rstick{\pket{0}} \qw
}\raisebox{-1.5em}{\hspace{10mm}.}
\end{equation}
Thus, up to a Gaussian operation on the output, we can see that the result of applying any operator $\op{O}$ before the first homodyne measurement in the macronode teleportation circuit is equivalent to applying $\op{O}$ before the homodyne measurement in the canonical cluster-state teleportation, but first squeezing and displacing the input state. For finite squeezing, the input state also undergoes $Q$-quadrature damping.

Suppose $\op{O}$ is the operator for photon subtraction of $n$ photons, $\op{\mathcal{S}}_n$, given previously by Eq.~\ref{eq:kraus_photsubt}. This is exactly the process described earlier in terms of the canonical cluster, and when followed by homodyne detection, can be written as an operator in $\op{Q}$. Using Eq.~\ref{eq:sub2fQ}, photon subtraction and homodyne detection becomes
\begin{equation}
    \pbra{m}\op{\mathcal{S}}_n=\pbra{0}f_n(\op{Q}),
\end{equation}
where $f_n(\op{x})$ is the same function derived previously, and is given by Eq.~\ref{eq:func_f_fin} in the Appendix. 
It is now clear that the macronode equivalent of the canonical case Kraus operator is just a slight modification of Eq.~\ref{eq:PhANTM_kraus}, and is given by
\begin{equation}
\scalebox{0.9}{$    \op{K}_n^{mac}=\sqrt{\tfrac{2\pi^{1/2}}{s}}\normalsize\op{G}\op{R}(\tfrac{\pi}{2})\op{K}_n  \op{X}^\dag_1(m_2)\op{S}(\tfrac{1}{2}\ln{2})e^{-\frac{(\op{Q}-\sqrt{2}m_2)^2}{2s^2}}$}.
\end{equation}
For large squeezing and weak subtraction beamsplitter reflectivity, we can use Eq.~\ref{eq:limit_case_final} and this becomes
\begin{align}
    &\op{K}_n^{mac}\propto \nonumber\\
    &\op{R}(\tfrac\pi2)\op{S}^\dag\left(\scriptstyle{\ln(\tfrac{1}{\sqrt{2}}\tanh{2\sq_0})}\right)\op{Z}^\dag(\tfrac{m_1}{\sqrt{2}})H_n(i\op{Q}-\tfrac{m_1+im_2}{\sqrt{2}}).
\end{align}
Just as before, subtracting photons before the homodyne measurement will apply a polynomial in $\op{Q}$ to the teleported quantum information. Here, however, the measurement dependent shift has both real and imaginary components. Additionally, there is a residual squeezing term due to the use of beamsplitters as the entangling gates in the macronode implementation from Eq.~\ref{eq:macro_gen_telep}. Failing to subtract any photons will only act to damp the state slightly and teleport it further along the cluster state, just as in the canonical case. 

An important point to note is that the subtraction must only be attempted on one wire of the macronode for the above derivation to hold. We have derived the results for applying the subtraction to the top wire before detection, but the derivation remains the same if the subtraction was instead performed on the middle wire, up to a rotation on the input. Because of this, there is no need to waste an intermediate step in applying an `empty' teleportation to enact a Fourier transform and reorient the state, as simply alternating which wire the subtraction is performed on will effectively apply the necessary rotation and allow for operators of the same quadrature to build up through several repetitions of the modified teleportation gadget.

\subsection{Effects of $m_2$}
\begin{figure}[h]
 \includegraphics[width = 0.45\textwidth]{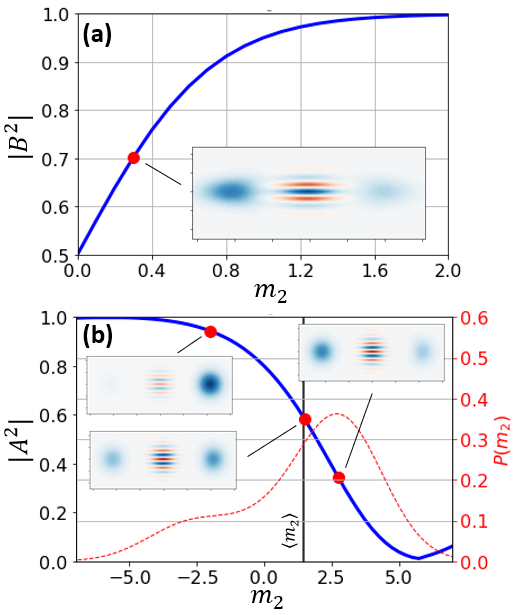}
  \caption{(a) The effects of the $Q$-basis homodyne measurement result ($m_2$ in Eq.~\ref{eq:macro_dict}) on the middle wire in the macronode teleportation scheme with photon subtraction occurring before the $P$-basis measurement. As $m_2$ increases, the negative displacement in the resulting cat-state superposition becomes more heavily weighted and $B$ increases. (b) Sending a weighted cat state into one round of PhANTM tends to reduce the more strongly weighted component. Here the initial state is set with $A^2=\tfrac{4}{5}$, $\alpha=4$, and $s_1=0$ while the cluster state is squeezed by $15$ dB. The weighting coefficient of the future cat, $A'$, tends to decrease.} 
    \label{fig:weighted_cats}
\end{figure}

We have shown how the PhANTM protocol can be translated from the canonical cluster state to a macronode implementation using the dictionary protocol, but we will leave an in-depth analysis on cat-state breeding and generation to future work. However, here we motivate the main differences created by the measurement-induced $Q$- quadrature shift by $m_2$. Sec.~\ref{sec:PhANTM} demonstrated that the overall effect of $m_1$ is to shift the phase between the components of the resulting cat state. Here, consider the case where $m_1=0$, so that the measurement results of $m_2$ can be dealt with separately. Instead of making a cat state that is a balanced superposition of coherent states  as per Eq.~\ref{eq:cat}, $H_n(i\op{Q}-\tfrac{im_2}{\sqrt{2}})$ applied to squeezed vacuum will result in the weighted superposition
\begin{equation}
    \ket{cat}_{A,B}\propto\left(A\op{X}(\alpha)\pm B\op{X}^\dag(\alpha)\right)S(\sq)\ketsub{0}{N},
    \label{eq:w_cat}
\end{equation}
where $A^2+B^2=1$ are the coefficients weighting the different components in the superposition.
When $m_2=0$, we have the case illustrated previously in Figs.~\ref{fig:QvsHerm} and~\ref{fig:cat_wig}, which corresponds to weighting coefficients of $A^2=B^2=\tfrac{1}{2}$. For nonzero $m_2$, the coefficients become unbalanced. This is shown in Fig.~\ref{fig:weighted_cats}(a) for the case of applying $H_4(i\sqrt{2}\op{Q}-im_2)$ to vacuum squeezed by 6 dB. The fourth-order Hermite polynomial was chosen since $\op{Q}^4$ and higher polynomials applied to squeezed vacuum become nearly indistinguishable from cat states in the idealized case, as illustrated by Fig.~\ref{fig:cat_wig}. For each value of $m_2$, a weighted cat state of the form of Eq.~\ref{eq:w_cat} was selected and numerically optimized to fit the resultant state, and we plot the fitted value of $B^2$ against $m_2$. Each optimized weighted cat had fidelity greater than $0.99$ with the resultant state, and it can be seen from the figure that the value of $B$ monotonically increases as $m_2$ increases.

Once the state becomes unbalanced, one can ask what happens when it is sent into future rounds of the PhANTM algorithm. To explore this, we consider input states to the circuit in Eq.~\ref{eq:macro_dict} that are small weighted cat states of the form of Eq.~\ref{eq:w_cat} with coefficients $A$ and $B$, and calculate the probability $P(m_2)$ of performing a homodyne measurement in mode two and obtaining outcome $m_2$. This is derived in Appx.~\ref{Appx:macro_Hprob} and given by
\begin{align}
P(m_2)=C\Big(A^2 &e^{-\frac{(\alpha-\sqrt{2} m_2)^{2}}{s_1^{2}+s_2^2}}+B^2 e^{-\frac{(\alpha+\sqrt{2} m_2)^{2}}{s_1^{2}+s_2^2}}\nonumber \\
&+2ABe^{-\tfrac{\alpha^2}{s_1^2}-\tfrac{2m_2^2}{s_1^2+s_2^2}}\Big),
\end{align}
where $s_1=e^{\sq_1}$ is the squeezing in the input weighted cat state, $s_2=e^{\sq_2}$ is the squeezing in each physical mode of the macronode cluster state, and the coefficient is
\begin{equation}
C=\left( \sqrt{\tfrac{\pi}{2}}\sqrt{s_1^2+s_2^2}\left(1+A B e^{-\frac{ \alpha^{2}}{ s_1^{2}}}\right)\right)^{-1}.
\end{equation}
This distribution is just two Gaussians originating from the input weighted cat state with broadening dependent on both the squeezing of the input state and the squeezing in the macronode cluster. The expectation value for a $Q$-basis homodyne measurement of mode two is given by
\begin{align}
\label{eq:exp_m2}
    \langle{\op{Q}_2}\rangle&=\int^\infty_{-\infty}\!dm_2 P(m_2)m_2 \nonumber \\
    &=\frac{\alpha(2A^2-1)}{\sqrt{2}(1+ABe^{-\alpha^2/s^2_1})},
\end{align}
which for $\alpha>0$, is always positive when $A^2>\tfrac{1}{2}$. This indicates that for an input cat state weighted more strongly toward the positive displacement ($\alpha>0$, $A>0$), the measurement result $m_2$ is likely to be positive, and thus the current round of the PhANTM algorithm will tend to increase the value of $B'$, the coefficient of the future cat state, and re-balance the weighting in the coherent state superposition. 

The re-balancing can be thought of as a `restoring force' tending to prevent either coherent state term from dominating the other, and is shown for a particular example in Fig.~\ref{fig:weighted_cats}(b), where an unbalanced cat state with amplitude $\alpha=4$ and initial weighting coefficient $A^2=\tfrac{4}{5}$ is sent through a macronode teleportation circuit with a single photon subtraction occurring. The measurement result $m_2$ is varied while $m_1$ is fixed to be zero to highlight the effects of changing $m_2$, and the output state is then numerically optimized to a fidelity above 0.99 with a weighted cat of the form of Eq.~\ref{eq:w_cat}. The blue curve plots the new value of the weighting coefficient, $A'^2$, against the measurement result where the probability distribution is shown as a red-dashed curve with the expectation value given by the solid vertical line. The figure insets display the Wigner functions for the output state and demonstrate that as $m_2$ increases, the weighting shifts from favoring the positive displacement ($A'>B'$) to favoring the negative displacement ($A'<B'$). As shown by the figure, the value of $A'$ is likely to decrease from the initial value of $A$. If the weighting remains unbalanced after the teleportation, then the process can be repeated until the output state has displacement components with weightings that are within a predetermined acceptable range.

A salient point of this process is that runaway effects will not occur; as shown by Eq.~\ref{eq:exp_m2}, the measurement result $m_2$ will be such that unbalancing is, on average, not exacerbated in the same direction. Overshooting is possible, but a strongly weighted coefficient will not progressively become larger until only one displacement remains in the superposition.  

\section{Experimental Imperfections}
\label{sec:exp_err}
While we have considered Gaussian noise due to finite squeezing, we have up to this point neglected other experimental imperfections that may be present, such as photon loss, detector inefficiency, excess antisqueezing, and displacement errors.  The effect of loss on state preparation is intolerable if it leads to squeezing below the fault-tolerance threshold~\cite{Menicucci2014ft,Fukui2018}. Therefore, in this section we will assume squeezing is high enough to reach the fault-tolerance threshold and and focus on the hitherto unknown effects on the protocol of excess noise and displacement errors.
\subsection{Effect of Detector Loss and Excess Noise}
Detectors with quantum efficiency $\eta<1$ can be modeled by placing a loss channel immediately prior to a detector with unit efficiency. Similarly, since loss to a squeezed state degrades the antisqueezing less than squeezing~\cite{Caves1981}, one can model a cluster state with excess antisqueezing as subjecting each single-mode squeezed vacuum to a loss channel before applying the entangling $\opCZ$ gates. Excess antisqueezing in cluster states has been previously shown not to impact the fault-tolerance threshold for a GKP encoding~\cite{Walshe2019}, but effects may be different for photon subtraction where photon-number resolving measurements are performed. \\
\indent Because both antisqueezing and detector imperfections can be modeled as losses, we use a simplified noise model for this work where we examine the effects of a photon-loss channel placed at different locations in the PhANTM circuit. This accounts for many of the expected real-world limitations, but Ref.~\cite{hillmann2021performance} considers more general imperfect measurements on bosonic encodings where loss and dephasing are both considered prior to detection. \\
\indent A loss channel can be modeled as placing a beamsplitter in the path of the quantum signal and tracing over the reflected mode such that the beamsplitter transmission coefficient is related to the channel efficiency, $\eta$, by $t^2=\eta$. Subjecting an initial quantum state $\rho$ to loss leads to a mixture given by
\begin{equation}
    \rho'=\sum^\infty_{l=0}\op{L}_l\rho\op{L}_l^\dag,
\end{equation}
where each $\op{L}_l$ is the Kraus operator for mode $a$ traveling through a loss channel given by
\begin{equation}
\label{eq:kraus_loss}
    \op{L}_l=\sqrt{\frac{(1-\eta)^l}{l!\eta^l}}\,\op{a}^l\,e^{\tfrac12\op{a}^\dag\op{a}\ln\eta}.
\end{equation}
\begin{figure}[h]
 \includegraphics[width = 0.45\textwidth]{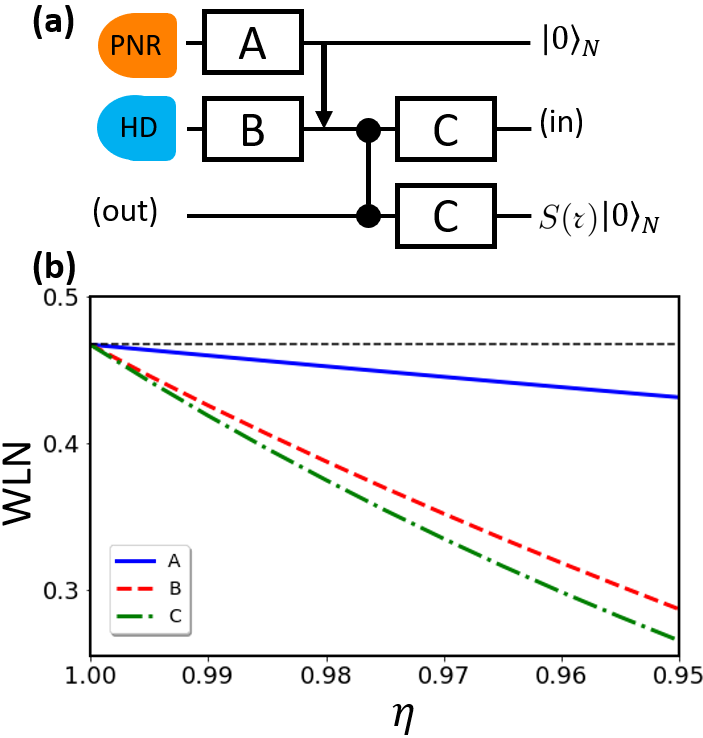}
  \caption{(a) PhANTM step measurement where a loss channel of efficiency $\eta$ is included before PNR detection (A), before homodyne detection (B), or immediately after squeezing before the $\opCZ$ gate is applied (C).  (b) Wigner logarithmic negativity (WLN) for a single PhANTM step post-selected on an $n=2$ photon subtraction and an $m=0$ homodyne detection with 11 dB of squeezing in the cluster state where losses occur at either A, B, or C locations. The black dashed line indicates the WLN for the ideal (lossless) case.} 
    \label{fig:loss_onestep}
\end{figure}
We model the imperfections by placing a loss channel at three locations of the PhANTM circuit as shown in Fig.~\ref{fig:loss_onestep}(a). Loss channels with efficiency $\eta$ at locations A and B model the respective imperfect PNR and homodyne detections, while losses at C account for excess antisqueezing. Because one would experimentally measure a value for squeezing after losses, the simulations here considered a more strongly squeezed initial state such that the measured squeezing would be held constant as the losses are increased. For example, simulations in this section were performed with squeezed states with 11 dB noise reduction compared to vacuum in the squeezed quadrature whereas the stretched quadrature was antisqueezed by more than 11 dB depending on the given value of loss.\\
\indent To measure the effects of loss, we employ the Wigner logarithmic negativity (WLN) to quantify the remaining quantum non-Gaussianity~\cite{albarelli2018resource}. This is defined as
\begin{equation}
    \text{WLN}(\rho)=\log{\left(\iint|W_\rho(q,p)|dq\,dp\right)},
\end{equation}
where the integral runs over all of phase-space. The WLN is useful as a metric to compare the relative quality of a class of states prepared with and without losses. Since we are not concerned with the particular state parameters within the family of cat states, such as the cat-state phase, for instance, WLN is more informative than fidelity for this particular application. Additionally, the WLN will decay to zero when nonclassical features vanish.

\begin{figure}[h]
 \includegraphics[width = 0.5\textwidth]{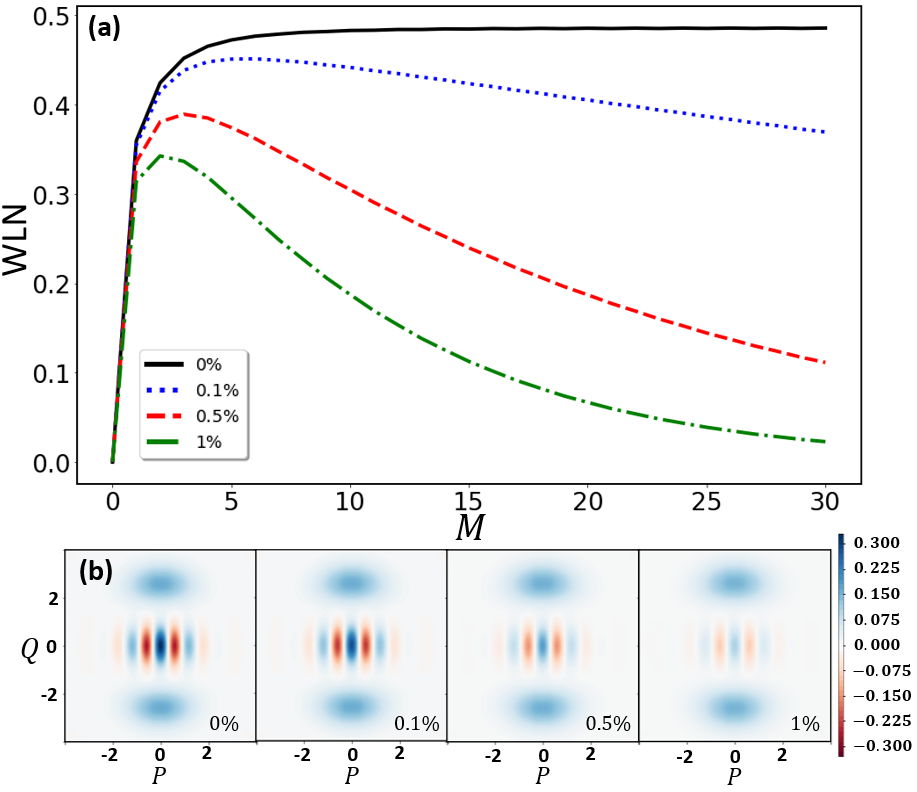}
  \caption{(a) WLN of the embedded state after $M$ PhANTM steps where losses occur after squeezed state preparation, before homodyne detection, and before PNR detection. Curves are shown for no loss and losses of $0.1\%,\, 0.5\%$, and $1\%$, and the indicated loss occurs at all locations (A, B, and C in Fig.~\ref{fig:loss_onestep}) at each PhANTM step, and at the same locations for the normal teleportation step to reorient the state before the next round of PhANTM. (b) Wigner functions of the embedded states after the $M=10$ PhANTM step for the given loss.} 
    \label{fig:loss_Mrounds}
\end{figure}

The WLN for a single round of PhANTM post-selected on a two-photon subtraction event and a homodyne detection of $m=0$ with loss at different locations is shown in Fig.~\ref{fig:loss_onestep}(b). As one would expect, the WLN decreases as losses increase. However, an interesting point to note is that the losses before PNR here are less detrimental, which can be attributed to the nature of photon subtraction. The reason for this becomes more evident on recalling the intuitive notion that for $\langle\op{N}\rangle$ average photons in a quantum state, the losses must be no more than $\sim \langle\op{N}\rangle^{-1}$ for quantum effects to persist and be useful, as has been proven for quantum estimation~\cite{Escher2011}. The relatively weak beamsplitter used for photon subtraction substantially reduces the mean photon number measured by the PNR detector; thus the threshold for tolerable losses will be higher. \\
\indent The utility of the PhANTM method is in its ability to be repeated on the cluster state to generate embedded cat states. Because of this, it is important to examine how losses compound and impact cat-state generation from one PhANTM step to another. This is shown in Fig.~\ref{fig:loss_Mrounds} where the WLN of the resultant states are shown after $M$ rounds of a PhANTM protocol that consists of photon-subtraction and teleportation followed by a single round of teleportation without subtraction as done previously in Sec.~\ref{subsec:catstates}. Here, however, equivalent losses ranging from $0.1\%$ to $1\%$ are included at all of the locations (A, B, and C) shown in Fig.~\ref{fig:loss_onestep}(a). This simulation assumes 11 dB of cluster state squeezing and post-selection of a single photon subtraction at each PhANTM step. Although the true photon subtractions would be stochastic, the mean number of photons subtracted at each step is $\langle n\rangle\sim 1$ (0.9 before the first subtraction, then $1<\langle n\rangle<1.15$ for remaining rounds), so assuming the mean is subtracted each time gives consistent results that are more comparable between different values of loss. Additionally, as stochastic homodyne measurements were shown previously to shift the cat state fringes, we also post-select on homodyne measurements of $m=0$ here to isolate the effects of loss.\\
\indent Fig.~\ref{fig:loss_Mrounds}(a) shows that the WLN begins at zero for an initial squeezed vacuum cluster state node and then begins to increase as photon subtractions de-Gaussify the embedded state. However, the WLN peaks and then begins to slowly decay as losses in the system degrade the state over repeated teleportations. This indicates that while cat states can be preserved against pure Gaussian noise as in Sec.~\ref{subsec:cat_stabilize}, the preservation does not extended to losses. Note that the zero loss case is also shown for comparision, and the WLN stabilizes at the point where Gaussian noise from finite squeezing balances the added non-Gaussianity introduced from photon subtraction at each step. \\
\indent While the losses reduce negativity, the location of the cat state interference fringes can still be discerned. This is shown in Fig.~\ref{fig:loss_Mrounds}(b), where the Wigner function for each output state after $M=10$ steps is plotted for the various loss levels. Provided that the losses are sufficiently low, a handful of PhANTM steps can still be performed to build up cat states embedded within the cluster state. When performed in parallel on several parts of the cluster state, this may still be sufficient to breed embedded GKP states, but an in-depth analysis of this type will be left to future work. 

One may have noticed that the levels of loss considered in this section are quite low, but this is not beyond the reach of expected experimental progress. Current transition-edge sensors have achieved photon-number-resolving capabilities with quantum efficiencies of $98 \%$~\cite{fukuda2011titanium}, and efficiencies above $99\%$ are possible~\cite{fujii2012thin}. Additionally, linear photodiodes as used in homodyne detection have reached measured quantum efficiencies nearing $100\%$ at telecommunications wavelengths~\cite{shen2022}.

\subsection{Displacement Errors}
In addition to losses, one must undo accrued displacements which can lead to experimental errors. As with the measurement-based CVQC model, one need not always physically undo the shifts due to stochastic homodyne measurements and may instead simply add an offset to future measurements results~\cite{Gu2009}. However, a physical displacement here will be required before each PNR measurement as one cannot directly project on a displaced Fock-basis eigenstate. This feed-forward displacement can be easily calculated by commuting previous homodyne results through the $\opCZ$ gate and subtraction beamsplitter, and then applied with an offline coherent state and beamsplitter with reflectivity $r<<1$~\cite{Paris1996}.\\
\indent Assume that the displacement operation has been imperfectly applied such that the results of the previous homodyne detections were undone up to some small residual $\Delta \alpha$. In this case, instead of the PNR detection of $n$ photons enacting a projection onto the Fock state $\ketsub{n}{N}$, it will instead project the state onto the displaced Fock state
\begin{align}
    \op{D}(\Delta \alpha)\ketsub{n}{N}\approx& \ketsub{n}{N}+\Delta \alpha\sqrt{n+1}\ketsub{n+1}{N} \nonumber\\
    &-\Delta \alpha^*\sqrt{n}\ketsub{n-1}{N}. 
\end{align}
\begin{figure}[h]
 \includegraphics[width = 0.5\textwidth]{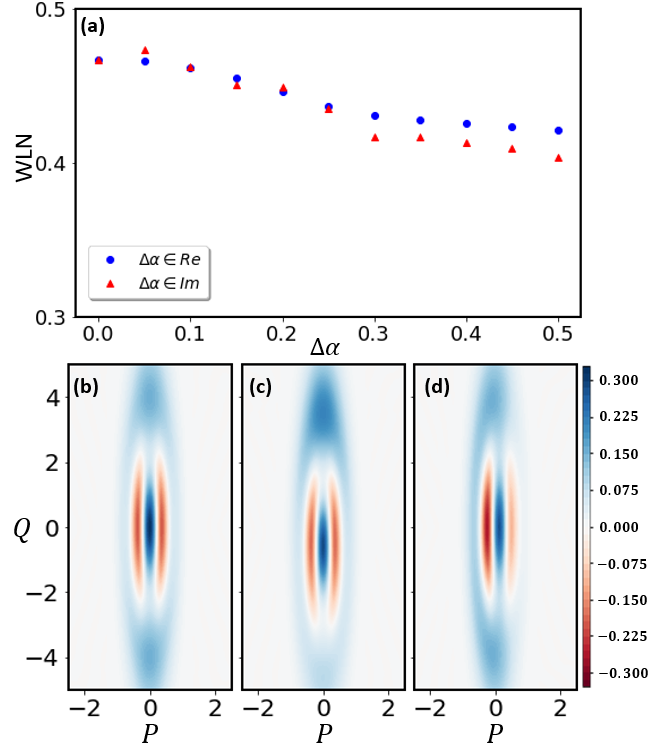}
  \caption{(a) WLN of the embedded state after a single PhANTM step post-selected on a 2 photon subtraction where there was a residual displacement error of $\Delta\alpha$ before PNR detection. Real and imaginary values of $\Delta\alpha$ were considered separately. (b) Wigner function of the embedded state with no error. (c) Wigner function for displacement error of $\Delta\alpha=0.2$ (d) and for displacement error of $\Delta\alpha=0.2i$.} 
    \label{fig:disp_err}
\end{figure}
The results of the imperfect displacements are shown in Fig.~\ref{fig:disp_err}, where real and imaginary values of $\Delta \alpha$ are considered separately. The simulation results are shown for a single PhANTM step post-selected on a PNR result of $n=2$, a homodyne detection result of $m=0$, and 11 dB squeezing in the cluster. In both the real and imaginary cases, the WLN only slightly decreases as $|\Delta\alpha|$ grows. The effects on the state can be more easily seen by examining the Wigner functions as shown in panels (b)-(d), where panel (b) shows the state with $\Delta\alpha=0$. For nonzero $\Delta \alpha$, the Wigner function becomes unbalanced depending on the phase of the displacement error. For purely real $\Delta \alpha$ as in panel (c), we see that the two coherent state-like lobes of the cat become unbalanced giving us a weighted cat state as was the case in Sec.~\ref{sec:macronode}. For purely imaginary $\Delta\alpha$ in panel (d), the interference fringes shift, which is exactly the same as the effects of varying homodyne measurement outcomes. For both panels (c) and (d), we chose $|\Delta\alpha|=0.2$. \\
\indent It should be noted that the values for displacement error shown are considerably higher than what may be achieved with current experimental techniques. When undoing the homodyne detection result, the displacement will be aimed along the imaginary axis, so the errors of fringe shifts shown in Fig.~\ref{fig:disp_err}(d) will come about due to imprecise amplitude control of a displacing field. However, state-of-the-art on-chip Mach-Zehnder interferometers allow for precise manipulation of amplitude control with up to -60 dB extinction~\cite{xie2022chip}. Real displacement errors leading to the weighting of the cat such as shown in Fig.~\ref{fig:disp_err}(c) will thus originate from instabilities in the phase of the displacing field. As far as this is concerned, current experiments have achieved phase noise below 2 mrad~\cite{Vahlbruch2016} leading to the conclusion that these errors can be virtually neglected.

\section{Conclusion}
\label{sec:conclusion}
Starting with a Gaussian cluster state, we have presented PhANTM, a method for using PNR detection and feed-forward Gaussian operations to locally de-Gaussify the cluster state to systematically generate Schr\"odinger cat states. Although each individual photon-subtraction event is probabilistic, teleporting quantum information along the cluster state and repeatedly applying steps of PhANTM leads to the production of cat states with high probability and mean amplitude dependent on the squeezing present in the cluster state. This process can be thought of as an adiabatic `cooling' toward a cat-state basis, as additionally seen by the ability for PhANTM to preserve embedded cat states as they are teleported throughout the cluster, preventing the build-up of excess Gaussian noise. The phase of the teleported cat will be randomized by homodyne measurement, but this randomization can be tracked by recording the measurement results at each step and fixed by performing a feed-forward displacement to align the cat fringe. This therefore allows a cluster state with PNR measurement capabilities to be used as a way to perpetuate a particular class of non-Gaussianity (cat states), without the need for GKP error-correction.\\
\indent Portions of 1-D quantum wires can each be converted into a cat state by repeatedly applying PhANTM to embed multiple cat states within a large cluster state, making this process compatible with state-of-the-art massively scalable 2D cluster states~\cite{Asavanant2019,Larsen2019}. Additionally, the photon-subtraction style measurement only requires low photon-number resolution, which is well within the current experimental capabilities of both TES systems~\cite{Lita2008,Morais2020,Eaton2022} and number-resolving silicon nanowire detectors~\cite{Cahall2017,endo2021quantum}.
Note that semiconductor detector technology, e.g.\ avalanche photodiodes, can be used in multiplexed detection schemes to achieve PNR~\cite{Fitch2003, Achilles2003,Nehra2020}, though this requires large numbers of multiplexed detectors since the individual detectors aren't PNR.

When one considers additional experimental imperfections, such as loss and imperfect displacements, the non-Gaussianity as measured by the WLN will begin to decay after too many PhANTM steps as losses compound. However, we find that PNR detection efficiency is less important than homodyne detection efficiency and that for sufficiently low loss, the WLN decays slowly. Fortunately, progress is being made in low-loss integrated photonic circuit platforms~\cite{liu2022ultralow} and in high quantum efficiency photodiodes~\cite{Zang2017}. Furthermore, photon subtraction has been experimentally demonstrated on existing cluster states~\cite{Ra2020}.

\indent While error-correction and universal QC with cat states alone is possible~\cite{Ralph2003, hastrup2021all}, we show how breeding protocols are compatible with cat states embedded within the cluster state solely by performing Gaussian measurements and feed-forward displacements locally on the cluster. This eliminates the need for offline resource-state generation as in current CV one-way error-corrected QC proposals~\cite{Bourassa2021blueprintscalable,larsen2021fault}, and allows for cat-state enlargement and GKP state synthesis. Taken together with the PhANTM cat-state generation protocol, we have provided a means to take Gaussian cluster states and transform them into universal quantum-computational resources by performing local homodyne detections, PNR measurements, and feed-forward displacements. 

\section*{Acknowledgments}
This work was supported by the Jefferson Lab LDRD project No. LDRD21-17 under which Jefferson Science Associates, LLC, manages and operates Jefferson Lab. The authors acknowledge Research Computing at The University of Virginia for providing computational resources and technical support that have contributed to the results reported within this publication. ME acknowledges support from the Metropolitan Washington Chapter ARCS Foundation Scholarship and wishes to thank CH Chang and A Hossameldin for helpful discussions.

\section{Appendix}\label{sec:appendix}
\subsection{The Damping Operator}
\label{appx:damp}
The damping operator is a non-unitary operator that acts to symmetrically de-amplify quantum states and can be defined as
\begin{equation}
    \hat{N}(\beta):=e^{-\beta \hat{a}^\dag \hat{a}}=e^{-\tfrac\beta2(\op{Q}^2+ \op{P}^2 - 1)}.
\end{equation}
When applied to a zero quadrature eigenstate, the damping operator brings the unphysical state to a finitely squeezed vacuum state.  This can be seen by applying $\hat{N}$ to the state $\pket{0}$. Writing $\pket{0}$ in the Fock-basis in the infinite squeezing limit, we have
\begin{align}
    \pket{0}&=\lim_{r\to\infty}\frac{1}{\sqrt{\cosh{r}}}\sum_{n=0}^\infty(\tanh{r})^n\frac{\sqrt{(2n)!}}{2^nn!}\ketsub{2n}{N}
\end{align}
Applying the damping operator to the above state yields
\begin{equation}
\resizebox{1\columnwidth}{!}{
\begin{math}
    \begin{aligned}
    \hat{N}&(\beta)\pket{0} =\\
    &\lim_{r\to\infty}\frac{1}{\sqrt{\cosh{r}}}\sum_{n=0}^\infty(e^{-2\beta}\tanh{r})^n\frac{\sqrt{(2n)!}}{2^nn!}\ketsub{2n}{N}.
    \end{aligned}
\end{math}
}
\end{equation}
By replacing $e^{-2\beta}\tanh{r}$ with $\tanh{r'}$, we have that
\begin{equation}
    \hat{N}(\beta)\pket{0}=\mathcal{N}_r\sum_{n=0}^\infty(\tanh{r'})^n\frac{\sqrt{(2n)!}}{2^nn!}\ketsub{2n}{N},
\end{equation}
which is just a finitely squeezed state of squeezing parameter $r'=\tanh^{-1}\left[{e^{-2\beta}}\right]$ in the limit $r\rightarrow \infty$ with normalization given by
\begin{equation}
    \mathcal{N}_r=\sqrt{\frac{\cosh{r}}{\cosh{r'}}}.
\end{equation}
Using these results, we can also examine the effects of applying the damping operator to a non-zero quadrature eigenstate, in which case we have
\begin{align}
    \hat{N}(\beta)\pket{m}&=\hat{N}(\beta)\hat{Z}(m)\hat{N}(-\beta)\hat{N}(\beta)\pket{0} \nonumber \\
    &=\hat{N}(\beta)\hat{Z}(m)\hat{N}(-\beta)\hat{S}(r')\ketsub{0}{N},
\end{align}
since the inverse of $\hat{N}(\beta)$ is $\hat{N}(-\beta)$. Using commutation relations, we can derive that
\begin{align}
\label{eq:damp_commute_a}
    \hat{N}(\beta)\hat{a}\hat{N}(-\beta)&= \hat{a} e^\beta\\
    \hat{N}(\beta)\hat{a}^\dag\hat{N}(-\beta)&= \hat{a}^\dag e^{-\beta}
\end{align}
which leads to 
\begin{equation} \label{eq:Qdamp}
   \hat{N}(\beta)\op{Q}\hat{N}(-\beta)= \op{Q}\cosh{\beta}+i\op{P}\sinh{\beta}.
\end{equation}
The above equation is useful to determine the transformation of the momentum-shift operator to be
\begin{align}
     \hat{N}&(\beta)\hat{Z}(m)\hat{N}(-\beta)=e^{im(\op{Q}\cosh{\beta}+i\op{P}\sinh{\beta})} \\
     &=e^{\frac{m^2}{2}\sinh{\beta}\cosh{\beta}}e^{-m(\op{P}\sinh{\beta})}\hat{Z}(m\cosh{\beta}).
\end{align}
Putting everything together, we now have that
\begin{equation}
\label{eq:damped_pstate}
    \hat{N}(\beta)\pket{m}=\mathcal{A}_\beta e^{-m(\op{P}\sinh{\beta})}\hat{Z}(m\cosh{\beta})\hat{S}(r')\ketsub{0}{N}
\end{equation}
where
\begin{equation}
\label{eq:appx_dampsq_coef}
    \mathcal{A}_\beta=e^{\frac{m^2}{2}\sinh{\beta}\cosh{\beta}} \mathcal{N}_r^{-1}.
\end{equation}

\subsection{Photon Subtraction}\label{Appx:phot_subt}
Direct application of the annihilation operator in the optical domain remains a challenge, but we have the capability to probabilistically apply $\op{a}^n$ followed by damping of both quadratures. This is implemented by a beamsplitter and PNRD as shown below, where a small amount of one mode is coupled to vacuum and sent to the PNRD. Detecting $n$ photons leads to the approximate application of $\op{a}^n$ when beamsplitter reflectivity is low, but we will derive the Kraus operator representation of this system in general. 
\begin{figure}[H]
\centering
 \includegraphics[width = 0.35\textwidth]{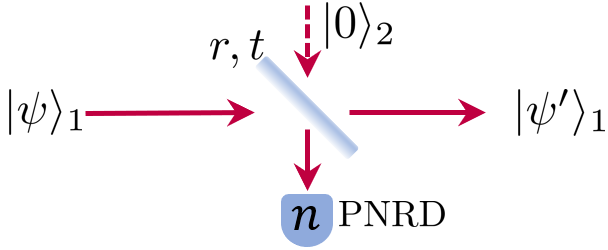}
  %\caption{}
    \label{fig:bsphot_sub}
\end{figure}
For a beamsplitter operator given by
\begin{equation}
    \bsop=e^{\theta(\op{a}_1^\dag \op{a}_2-\op{a}_1\op{a}_2^\dag)},
\end{equation}
the annihilation and creation operators are transformed to
\begin{equation}
    \bsop^\dag \begin{pmatrix}
\op{a}_1\\
\op{a}_2\end{pmatrix}\bsop =
\begin{pmatrix}
t\op{a}_1+r\op{a}_2\\
t\op{a}_2-r\op{a}_1\end{pmatrix},
\end{equation}
where $r=\sin{\theta}$ and $t=\cos{\theta}$. If we write $\ket{\psi}=\sum_m \psi_m \ket{m}$, then coupling $\ket{\psi}$ to vacuum with $\bsop$ and detecting $n$ photons in one output leads to the final state
\begin{equation}
\resizebox{1\columnwidth}{!}{
\begin{math}
    \begin{aligned}
        \ket{\psi'}&=\brasub{n}{2} \sum_{m=0}\frac{\psi_m}{\sqrt{m!}} \bsop\op{a}_1^{\dag m}\ketsub{0}{1}\ketsub{0}{2}  \\
        &=\brasub{n}{2} \sum_{m=0}\frac{\psi_m}{\sqrt{m!}} \sum_{k=0}^m \binom{m}{k}(t\op{a}_1^{\dag})^{m-k}(-r\op{a}_2^\dag)^k\ketsub{0}{1}\ketsub{0}{2} \\
        &=\frac{1}{\sqrt{n!}}\left(\frac{-r}{t}\right)^n \sum_{m=0}t^m\psi_m \op{a}_1^n\ketsub{m}{1}  \\
    \end{aligned}
\end{math}
}
\end{equation}

Recognizing that $t^m\ket{m}$ can be rewritten as an exponentiated number operator acting on a Fock state, we can replace $t^m$ with a damping operator having argument $(-\ln{t})$ and arrive at
\begin{equation}
    \ket{\psi'}=\frac{1}{\sqrt{n!}}\left(\frac{-r}{t}\right)^n \op{a}^n \op{N}(\beta)\ket{\psi},
\end{equation}
where $\beta=-\ln{t}$. Using the commutation relation from Eq.~\ref{eq:damp_commute_a} allows us to write the Kraus operator representing photon-subtraction for any beamsplitter reflectivity as
\begin{equation}
\op{\mathcal{S}}_n=\frac{(-1)^ne^{-n\beta/2}}{\sqrt{n!}}\left(2\sinh{\beta}\right)^{n/2}\hat{N}(\beta)\hat{a}^n.
    \label{eq:appxkraus_photsubt}
\end{equation}
The probability of successfully subtracting $n$ photons from an input density matrix, $\rho$, is given by 
\begin{equation}
    P(n)=\text{Tr}\left[\op{\mathcal{S}}^\dag_n\op{\mathcal{S}}_n \rho  \right],
\end{equation}
and the new subtracted density matrix becomes
\begin{equation}
\rho'=  \frac{\op{\mathcal{S}}_n \rho \op{\mathcal{S}}^\dag_n }{P(n)}. 
\end{equation}

\subsection{PhANTM Derivation}\label{Appx:main_derivation}
We fully derive the effects of photon subtraction proceeded by teleportation in the presence of finite squeezing. First, we derive the circuit identity given by the following lemma:
\begin{lemma}
\begin{equation}  \label{eq:subt_cir_id}
       \begin{split}
       \centering
 \Qcircuit @C=0.3cm @R=0.8cm {
     \lstick{\brasub{n}{N}} &\bsbal{1} &\qw& \rstick{\ketsub{0}{N}}&&& \\
     \lstick{\brasub{m}{P}}&\qw& \qw &&}{\raisebox{-1.3em}{$=$}}\hspace{1em}
     \Qcircuit @C=0.2cm @R=0.1cm {\vspace{2em}&&&&\\&&&&\lstick{\brasub{0}{P}}&\gate{f_n(Q)}&\qw
    }\hspace{4mm}{\raisebox{-.8em}{,}} 
       \end{split}\, 
\end{equation}
\end{lemma}
where the arrow in the diagram represents a beamsplitter, $\bsop_\theta$, with reflectivity $r=\sin{\theta}$ and transmissivity $t=\cos{\theta}$. 
The lefthand side can be written as
\begin{equation}
\label{eq:sub_cir_ops}
    \brasub{n}{N_1}\brasub{m}{P_2}\bsop_\theta\ketsub{0}{N}\otimes\id,
\end{equation}
and in the $Q$-basis, the bras can be expressed as 
\begin{align}
    \brasub{n}{N}&=\int\! du \qbra{u}\,\psi_n\!(u)\\
    \brasub{m}{P}&=\frac{1}{\sqrt{2\pi}}\int\! dv \qbra{v}\,e^{-imv},
\end{align}
where $\psi_n\!(x)$ is the wavefunction of the harmonic oscillator:
\begin{equation}
    \psi_n\!(x)=\frac{1}{\pi^{1/4}\sqrt{2^nn!}}e^{-x^2/2}H_n(x).
\end{equation}
Using the above expressions, and temporarily neglecting the $(2\pi)^{-1/2}$ factor, Eq.~\ref{eq:sub_cir_ops} becomes
\begin{align}
    &\int\!dudv\brasub{u}{Q_1}\brasub{v}{Q_2}\psi_n\!(u)e^{-imv}\bsop_\theta \ketsub{0}{N_1}\\
    =&\text{\footnotesize$\int\!dudv\brasub{0}{Q_1}\brasub{0}{Q_2}\bsop_\theta\bsop^\dag_\theta e^{iu\op{P}_1+iv\op{P}_2}\psi_n\!(u)e^{-imv}\bsop_\theta \ketsub{0}{N_1}$}\\\
    =&\text{\footnotesize$\int\!dudv\brasub{0}{Q_1}\brasub{0}{Q_2} e^{iu(t\op{P}_1-r\op{P}_2)+iv(t\op{P}_2+r\op{P}_1)}\psi_n\!(u)e^{-imv}\ketsub{0}{N_1}$}
\end{align}
where in the last line we have used the fact that a beamsplitter applied to a pair of $q=0$ states has no effect. Expressing the vacuum state in the $Q$-basis as well yields
\begin{align}
    &\int\!dudvdz\brasub{tu+rv}{Q_1}z\rangle_{Q_1} \brasub{tv-ru}{Q_2}\psi_n\!(u)\psi_0\!(z)e^{-imv}\\
    =&\int\!dudv\brasub{tv-ru}{Q_2}\psi_n\!(u)\psi_0\!(tu+rv)e^{-imv}\\
    =&\frac{1}{\sqrt{2\pi}}\int\!dx\brasub{x}{Q}f_n(x)\\
    =&\pbra{0} f_n(\op{Q})
\end{align}
where we reintroduced the missing prefactor in the final two lines. With a change of the integration variables to $x=tv-ru, y=tu+rv$, one can see that
\begin{equation}
    f_n(x)=\int\!dy\, \psi_n(ty-rx)\psi_0(y)e^{-im(tx+ry)}.
\end{equation}
Changing variables again, this can be rewritten as
\begin{align}
\label{eq:func_f}
    f_n(x)&=\frac{1}{t}\int\!dy\, \psi_n(y)\psi_0(\frac{y+rx}{t})e^{-\tfrac{im}{t}(ry+x)}\nonumber\\
    &=\tfrac{e^{-\sq/2}}{t}e^{-i\zeta x }\int\! dy\,\brasub{n}{N}y\rangle_Q\, {}_Q\!\langle y| \op{D}(\alpha)\op{S}(\sq) \ketsub{0}{N}\nonumber\\
    &=\tfrac{e^{-\sq/2}}{t}e^{-i\zeta x}\bra{n}\op{D}(\alpha)\op{S}(\sq) \ket{0}
    %&=\tfrac{e^{\sq/2}\pi^{1/4}}{t\sqrt{\cosh{\sq}}}e^{-i\zeta x}\sum^\infty_{k=0}(\tanh{\sq})^k\frac{\sqrt{(2k)!}}{2^k k!}\bra{n}\op{D}(\alpha)\ket{2k},
\end{align}
where here, $\alpha=-2^{-1/2}(rx+i\tfrac{r}{t}m)$, $\sq=\ln{t}$, and $\zeta=\tfrac{m}{t}(1+\tfrac12 t^2)$.

At this point, it becomes necessary to derive an identity for the overlap coefficients of the Fock states with displaced, squeezed vacuum. This overlap can be written as
\begin{equation}
\scalebox{0.9}{
    $\bra{n}\op{D}(\alpha)\op{S}(\sq) \ket{0}=e^{-\tfrac{i}{2}\gamma\beta}\bra{n}\op{Z}(\gamma)\op{X}(\beta)\op{S}(\sq) \ket{0}$},
\end{equation}
where $\gamma=\sqrt{2}\text{Im}[\alpha]$ and $\beta=\sqrt{2}\text{Re}[\alpha]$. Transforming to the $Q$-basis, we have
\begin{align}
    &e^{-\tfrac{i}{2}\gamma\beta}\int\!dsdt\psi_n(s)\psi_0(t)\qbra{s}\op{Z}(\gamma)\op{X}(\beta)\op{S}(\sq)\qket{t}\nonumber \\
    =&e^{\tfrac{\sq}{2}-\tfrac{i}{2}\gamma\beta}\int\!dsdt\psi_n(s)\psi_0(t)e^{is\gamma}\delta(s-\beta-te^{-\sq})\nonumber \\
    =&\frac{e^{\tfrac{\sq}{2}-\tfrac{i}{2}\gamma\beta}}{\sqrt{2^n n! \pi}}\int\!dse^{-s^2/2}H_n(s)e^{-(s-\beta)^2e^{2\sq}/2}e^{is\gamma}\nonumber\\
    &=\frac{e^{\tfrac{\sq}{2}-i\gamma(\beta/2-B)+C}}{A\sqrt{2^n n! \pi}}\int\!dse^{-s^2/2}H_n(\tfrac{s}{A}+B)e^{i\tfrac{s}{A}\gamma},
    \label{eq:ovlp_derv_midstep}
\end{align}
where we have defined
\begin{align}
    A&=\sqrt{1+e^{2\sq}}\label{eq:A}\\
    B&=\frac{\beta e^{2\sq}}{1+e^{2\sq}}\label{eq:B}\\
    C&=\frac{\beta^2 e^{4\sq}}{2(1+e^{2\sq})}-\frac12\beta^2e^{2\sq}\label{eq:C}.
    \end{align}
We can now make use of the pair of Hermite polynomial identities
\begin{align}
     H_n(x+y) &= 2^{-\frac n 2}\sum_{k=0}^n \binom{n}{k} H_{n-k}\left(x\sqrt 2\right) H_k\left(y\sqrt 2\right)\\
     H_n(\gamma x) &= \sum_{i=0}^{\left\lfloor \tfrac{n}{2} \right\rfloor} \gamma^{n-2i}(\gamma^2 - 1)^i \binom{n}{2i} \frac{(2i)!}{i!} H_{n-2i}(x),
\end{align}
along with the property that Hermite polynomials are eigenfunctions of the Fourier transform, i.e.,
\begin{equation}
\scalebox{0.9}{$
    \int^{\infty}_{-\infty}\!dxe^{ikx}e^{-x^2/2}H_n(x)=\sqrt{2\pi}(i)^n e^{-\frac12 k^2} H_n(k)$},
\end{equation}
to take Eq.~\ref{eq:ovlp_derv_midstep} and arrive at the result of
\begin{align}
    %\tfrac{i^ne^{\tfrac{\sq}{2}-i\gamma(\beta/2-B)+C}}{A^{n+1}\sqrt{2^{n-1} n!}}e^{-\tfrac{\gamma^2}{2A^2}}\sum_{k=0}^n\sum^{\left\lfloor \tfrac{n-k}{2} \right\rfloor}_{j=0}\binom{n}{k}\binom{n-k}{2j}\frac{(2j!)A^{2j+k}}{j!\sqrt{-2^{2j+k}}}(\tfrac{2}{A^2}-1)^jH_k(\sqrt{2}B)H_{{n-k-2j}}(\tfrac{\gamma}{A})
    \bra{n}&\op{D}(\alpha)\op{S}(\sq) \ket{0}=\nonumber\\    
   &\kappa\sum_{k=0}^n\sum^{\left\lfloor \text{$\tfrac{n-k}{2}$} \right\rfloor}_{j=0}\chi_{j,k}H_k(\sqrt{2}B)H_{{n-k-2j}}(\tfrac{\gamma}{A})
   \label{eq:ovlp_dsqfock}
\end{align}
where
\begin{align}
    \kappa&= \frac{i^n e^{\tfrac{\sq}{2}-i\gamma(\beta/2-B)+C}}{A^{n+1}\sqrt{2^{n-1} n!}}e^{-\tfrac{\gamma^2}{2A^2}}\label{eq:kappa}\\
    \chi_{j,k}&=\binom{n}{k}\binom{n-k}{2j}\frac{(2j!)A^{2j+k}}{j!\sqrt{-2^{2j+k}}}(\tfrac{2}{A^2}-1)^j\label{eq:chi}.
\end{align}
We can now put everything back together and see that $f_n(x)$ is a polynomial of degree $n$ with an exponential envelope. Using Eq.~\ref{eq:func_f} along with the derived overlap in Eq.~\ref{eq:ovlp_dsqfock} and the intermediary definition of Eqs.~\cref{eq:A,eq:B,eq:C,eq:kappa,eq:chi} lead to
\begin{equation}
    f_n(x)=e^{-imx(\tfrac{2t}{1+t^2})}e^{-x^2\tfrac{t^2r^2}{2+2t^2}}f_n'(x),
    \label{eq:func_f_fin}
\end{equation}
where $f_n'(x)$ is the polynomial portion given by
\begin{widetext}
\begin{equation}
    f_n'(x)=\tfrac{i^n}{\sqrt{2^{n-1}n!(1+t^2)^{n+1}}}\sum_{k=0}^n\sum^{\left\lfloor \text{$\tfrac{n-k}{2}$} \right\rfloor}_{j=0}\binom{n}{k}\binom{n-k}{2j}\frac{(2j)!r^{2j}}{2^j j!}\left(\tfrac{1+t^2}{2}\right)^{k/2}(-i)^{k+2j}H_k(\tfrac{-x\sqrt{2}rt^2}{1+t^2})H_{{n-k-2j}}(\tfrac{-mr}{t\sqrt{1+t^2}})
   \label{eq:fprime_appx}
\end{equation}
\end{widetext}
Subtraction-assisted teleportation can thus be recognized as a circuit of the form

\begin{equation}\label{eq:}
    \Qcircuit @C=0.8em @R=1.5em {
\lstick{\brasub{0}{P}} &\qw&\gate{f_n(Q)} &\ctrl{1} & \qw& \rstick{(\text{in})} \qw \\
%\lstick{\brasub{\phi}{\text{out}}} &\qw&\qw&\control \qw & \gate{S(\mf{R'})} &\rstick{\ketsub{0}{N}} \qw
\lstick{(\text{out})}&\qw&\qw&\control \qw & \gate{S(\sq')} &\rstick{\ketsub{0}{N}} \qw
}\raisebox{-1.9em}{\hspace{10mm},}
\end{equation}
where $f$ is the resulting function from Eq.~\ref{eq:func_f_fin}. The operator in the top wire is a function of $\op{Q}$ only so commutes with the $\opCZ$, and we can make use of the fact that we can write the input momentum-squeezed state as
\begin{align}\label{eq:appx_sq_trick}
    \hat{S}(\sq')\ketsub{0}{N}&=\pi^{-1/4}\int dt  e^{-t^2/2s^2}\qket{t}\nonumber \\
    &=\pi^{-1/4}e^{-\op{Q}^2/2s^2}\int dt \qket{t} \nonumber \\
    &=\pi^{1/4}\sqrt{\frac{2}{s}}e^{-\op{Q}^2/2s^2}\pket{0},
\end{align}
where $s=e^{\sq'}$. In this form, the exponentiated function of $\op{Q}$ commutes through the $C_Z$ gate to give us
\begin{equation}
    \Qcircuit @C=0.8em @R=1.2em {
\lstick{\brasub{0}{P}} &\qw&\qw&\ctrl{1} & \gate{f_n(Q)} & \rstick{(\text{in})} \qw \\
%\lstick{\brasub{\phi}{\text{out}}} &\qw&\gate{e^{-Q^2/2s^2}}&\control \qw & \qw &\rstick{\pket{0}} \qw
\lstick{(\text{out})}&\qw&\gate{e^{-Q^2/2s^2}}&\control \qw & \qw &\rstick{\pket{0}} \qw
}\raisebox{-1.5em}{\hspace{10mm},}
\end{equation}
which is easily recognizable as the teleportation circuit with additional operators applied both before and after the teleportation. The overall Kraus operator representation of the action of the circuit acting on the input is given by
\begin{equation}
\label{eq:SAT_kraus_apx}
    \op{K}=\pi^{1/4}\sqrt{\frac{2}{s}}e^{-\op{Q}^2/2s^2}\op{R}(\tfrac{\pi}{2})f_n(\op{Q}).
\end{equation}
However, if we note that there is a $Q$-quadrature shift within $f_n(\op{Q})$ that we may wish to undo with feed-forward operations, we can commute this to the back of the Kraus operator. This leads to an alternate form of
\begin{align}
    \op{K}=&\tfrac{\pi^{1/4}\sqrt{2}}{\sqrt{s}}e^{\tfrac{m^2}{s^2}\sigma^2}\op{X}(m\sigma)\op{R}(\tfrac{\pi}{2})\nonumber \\
    &\times e^{-\tfrac{1}{2s^2}(\op{P}-m\sigma)^2}e^{-\op{Q}^2 \tfrac{\sigma t(1-t^2)}{4}}f_n'(\op{Q}),
\end{align}
where we define $\sigma=\tfrac{2t}{1+t^2}$. For weak beamsplitter reflectivity, we have that $t\rightarrow 1$ and $\sigma\rightarrow1$, and in this limit, we can recover the form of the idealized case discussed in the main text. Taking the large squeezing limit $s\rightarrow\infty$ and weak beamsplitter reflectivity $r\rightarrow0$, we have to leading order in $r$ that 
\begin{align}
    &f'_n(\op{Q})\rightarrow \sum^{\left\lfloor \text{$\tfrac{n}{2}$} \right\rfloor}_{j=0}\frac{\sqrt{n!}2^{-j}r^{2j}}{2^n j!(n-2j)!} \times \nonumber\\
    &\sum^{n-2j}_{k=0}\binom{n-2j}{k}H_k(\tfrac{-Qr}{\sqrt{2}})i^{n-k-2j}H_{n-k-2j}(\tfrac{-mr}{\sqrt{2}}).
    \label{eq:lim_case}
\end{align}

Defining the generalized probabilist's Hermite polynomial $H_{e_n}^{[\alpha]}(x)$ as
\begin{equation}
    H_{e_n}^{[\alpha]}(x) = \alpha^{\frac{n}{2}}H_{e_n}\left(\frac{x}{\sqrt{\alpha}}\right)=\left(\tfrac{\alpha}{2}\right)^{\frac{n}{2}}H_{n}\left(\frac{x}{\sqrt{2\alpha}}\right)
\end{equation}
and making using of the identity
\begin{equation}
    H_{e_n}^{[\alpha+\beta]}(x + y) = \sum_{k=0}^n \binom{n}{k} H_{e_k}^{[\alpha]}(x) H_{e_{n-k}}^{[\beta]}(y),
\end{equation}
along with the special case of
\begin{equation}
    H_{e_n}^{[0]}(x)=(x)^n
\end{equation}
allows us to simplify Eq.~\ref{eq:lim_case} to
\begin{align}
    f'_n(\op{Q})&\rightarrow \frac{r^n\sqrt{n!}}{2^n \sqrt{2^n}}\sum^{\left\lfloor \text{$\tfrac{n}{2}$} \right\rfloor}_{j=0}\frac{\left(\sqrt{2}(Q+im)\right)^{n-2j}}{j!(n-2j)!}.
\end{align}
We can now simplify the above expression by using the expansion of the Hermite polynomial given by
\begin{equation}
    H_n(x) = n! \sum_{m=0}^{\left\lfloor \tfrac{n}{2} \right\rfloor} \frac{(-1)^m}{m!(n - 2m)!} (2x)^{n - 2m},
\end{equation}
which leads to the limiting case Kraus operator of
\begin{equation}
    \op{K}\propto \op{X}(m)\op{R}(\tfrac{\pi}{2})H_n\left(\frac{i\op{Q}-m}{\sqrt{2}}\right).
    \label{eq:limit_case_final_appx}
\end{equation}

\subsection{Hermite Polynomial}\label{Appx:herm_proof}
We prove the relation
\begin{equation}
\label{eq:appx_herm}
    H_{e_{n}}(Q)=\sum_{k=0}^{n} \binom{n}{k}  (-i P)^{k}(Q+i P)^{n-k}.
\end{equation}
This is easily verified to be true for the first few terms, so we proceed by induction. Assuming Eq.~\ref{eq:appx_herm} is true for $n$, we can make use of the relation $\binom{n+1}{k} =\binom{n}{k} + \binom{n}{k-1}$ to prove the validity for $n+1$ as follows:
\begin{widetext}
\begin{gather}
	\sum_{k=0}^{n+1} \binom{n+1}{k}\left(- i P\right)^{k}(Q+i P)^{n+1-k}=\\
	\sum_{k=0}^{n} \binom{n}{k}\left(- i P\right)^{k}(Q+i P)^{n-k}(Q+i P)+\sum_{k=1}^{n} \binom{n}{k-1}\left(- i P\right)^{k}(Q+i P)^{n+1-k}+\left(-iP \right)^{n+1} \\	
	=H_{e_{n}}(Q)(Q+i P)+\sum_{k=0}^{n-1}\binom{n}{k}(-i P)^{k+1}(Q+i P)^{n-k}+(-i P)^{n+1}\\
	=H_{e_{n}}(q)(Q+i P)+(-i P)\left(\sum_{k=0}^{n}\binom{n}{k}(-i P)^{k}(Q+i P)^{n-k}\right)\\
	=H_{e_n}(Q)(Q+i P)+\left( -i P\right) H_{e_{n}}(Q)=Q H_{e_{n}}(Q)+\underbrace{\left[-i P, H_{e_{n}}(Q)\right]}_{-\frac{d H_{e_n}(Q)}{d Q}}\\
	=Q H_{e_{n}}(Q)-\frac{d H_{e_n}(Q)}{d Q}	\equiv H_{e_{n+1}}(Q)
\end{gather}	
\end{widetext}

QED.
\subsection{Example evolution}
\label{Appx:example}

\begin{figure}[h]
 \includegraphics[width = 0.49\textwidth]{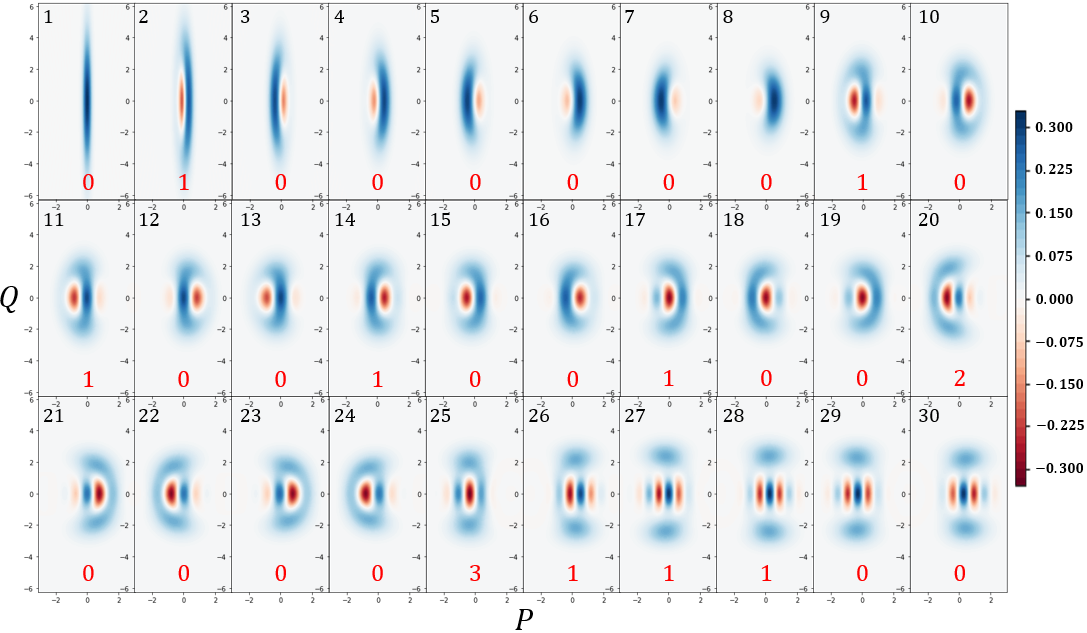}
  \caption{Typical results from iterated PhANTM where measurement results are stochastic. Each Wigner function plotted shows what state is now $\opCZ$'ed to the cluster after the $M$-th step (black numbering) where the number of subtraction photons from the corresponding step is shown in red.}
    \label{fig:example_PhANTM}
    \end{figure}
Fig.~\ref{fig:example_PhANTM} shows the full progression of the PhANTM algorithm as quantum information is teleported along the canonical cluster state with photon subtraction attempted before homodyne measurements on every-other teleportation. Steps without attempting photon subtraction are used for teleporting the state and applying a $\pi/2$ rotation to reorient the state. The Wigner function for the state at each step where subtraction was attempted is shown, with the number of actual photons subtracted listed in red. Each teleportation takes into account finite squeezing of 15 dB in the cluster state, randomly sampled homodyne and subtraction measurement outcomes, and a subtraction beamsplitter with reflectively chosen to have the same order damping effect as finite squeezing.

\subsection{Breeding on the Cluster State}\label{Appx:breed}
Grid state breeding with beamsplitters has been discussed in detail in Ref.~\cite{Weigand2018}. Here we provide the analogous derivation for cluster states. Suppose we have a superposition of evenly spaced Gaussian peaks, which can generally be described by a product of cat-like operators acting on squeezed vacuum as
\begin{equation}
\label{eq:appx_grid}
   \ket{\Psi_{{\alpha},\varphi,r}} \propto \mathcal{\op{D}}(\alpha,\bm{\varphi}) \op{S}(r) \ket{0}.
\end{equation}
where we define
\begin{equation}
    \mathcal{\op{D}}(\alpha,\bm{\varphi}):=\prod_k^{N}\left(\op{Z}(\alpha)+e^{i \varphi_{k}} \op{Z}^{\dag}(\alpha)\right).
\end{equation}
If one teleports through this state as opposed to the normal momentum squeezed state, then the circuit will appear as
\begin{equation}\label{circ:breed}
\centering
    \Qcircuit @C=1.2em @R=1.5em {
\lstick{\brasub{m}{P}} &\qw& \ctrl{1} & \rstick{\ket{\psi}_1} \qw \\
%\lstick{\bra{\phi}} &\qw&\control \qw &\rstick{\ketsub{\Psi{\alpha,\varphi,r}}{2}} \qw
&\qw&\control \qw &\rstick{\ketsub{\Psi{\alpha,\varphi,r}}{2}} \qw
}\raisebox{1em}{\hspace{16mm}}.
\end{equation}
Writing the squeezed vacuum of mode two as an operator in $\op{Q}$ applied to a momentum eigenstate as in Eq.~\ref{eq:sq_trick} and commuting all remaining operators on mode two through the $\opCZ$ reveals that the effect of this circuit is to apply the operator
\begin{equation}
    \op{\mathcal{G}}=\mathcal{\op{D}}(\alpha,\bm\varphi) e^{-\op{Q}^2/2s^2}\op{R}(\tfrac{\pi}{2})\op{Z}^\dag(m),
\end{equation}
where $s=e^r$. Suppose the input to this circuit was of the same form as Eq.~\ref{eq:appx_grid}, but first having undergone a Fourier transform. The transformed state would then be
\begin{align}
    \ket{\phi}&\propto\mathcal{\op{G}}\op{R}(\tfrac{\pi}{2})\ket{\Psi_{\alpha',\varphi',r'}}\nonumber \\
    &\propto\mathcal{\op{D}}(\alpha,\bm\varphi) e^{-\op{Q}^2/2s^2}\op{X}(m)\op{R}(\pi)\mathcal{\op{D}}(\alpha,\bm\varphi')\op{S}(r')\ket{0}.
\end{align}
Making use of the relations that
\begin{align}
\op{Z}(\alpha_k)\op{X}(\alpha_j)&=e^{i\alpha_k\alpha_j}\op{X}(\alpha_j)\op{Z}(\alpha_k),\\
    e^{-\op{Q}^2/2s^2}\op{X}(m)&=e^{m^2/2s^2}\op{X}(m)e^{-\tfrac{1}{s^2}\left(m\op{Q}+\op{Q}^2\right)},
\end{align}
we can write the evolved state as
\begin{equation}
\resizebox{1\hsize}{!}{$
    \ket{\phi}\propto\op{X}(m)\mathcal{\op{D}}(\alpha,\bm\varphi-2m\alpha)\mathcal{\op{D}}(-\alpha,\bm\varphi')e^{-\tfrac{1}{s^2}\left(m\op{Q}+\op{Q}^2\right)}\op{S}(r')\ket{0}$
    },
\end{equation}
where $\bm\varphi-2m\alpha$ indicates that each phase in the expanded product has been shifted by the measurement result and the value of $\alpha$ to be $\varphi_k\rightarrow \varphi_k-2m\alpha
$. With the help of Eq.~\ref{eq:sq_trick}, the exponentiated quadratic in $\op{Q}$ applied to squeezed vacuum can be rewritten as 
\begin{align}
    e^{-\tfrac{1}{s^2}\left(m\op{Q}+\op{Q}^2\right)}\op{S}(r')\ket{0}&=e^{-\tfrac{1}{s^2}\left(m\op{Q}+\op{Q}^2\right)-\tfrac{1}{s'^2}\op{Q^2}}\ketsub{0}{P} \nonumber \\
    &=\op{X}(B)e^{-A\op{Q}^2}\op{X}^\dag(B)\pket{0}\nonumber \\
     &=\op{X}(B)e^{-A\op{Q}^2}\pket{0}
\end{align}
where $s'=e^{r'}$ and
\begin{align}
    A&=\frac{1}{2}(\frac{1}{s^2}+\frac{1}{s'^2})\\
    B&=\frac{ms^2s'}{s^2+s'^2}.
\end{align}
Commuting the obtained position-quadrature shift to the left and converting the exponentiated $\op{Q}^2$ back to a squeezer applied to vacuum, we have
\begin{align}
    \ket{\phi}\propto&\op{X}(m+B)\mathcal{\op{D}}(\alpha,\bm\varphi-2(m+B)\alpha)\nonumber \\
    &\times \mathcal{\op{D}}(-\alpha,\bm\varphi'+2B\alpha)\op{S}(r'')\ket{0} \nonumber \\
    \propto& \op{X}(m+B)\mathcal{\op{D}}(\alpha,\bm\varphi'')\op{S}(r'')\ket{0},
\end{align}
which is exactly a state of the form of Eq.~\ref{eq:appx_grid} with an additional displacement and modified squeezing
\begin{equation}
    r''=r+r'-\frac{1}{2}\ln{\left(e^{2r}+e^{2r'}\right)}
\end{equation}

\subsection{Beamsplitter to $\opCZ$ Breeding}
\label{Appx:BS2CZ}
Several other works have discussed breeding cat states for enlargement or to make grid states in the context of beamsplitters followed by a homodyne measurement. Here, we show how this is translated to the canonical cluster state measurements through the use of the beamsplitter unitary decomposition. Breeding cat states is achieved by using the circuit
\begin{equation}\label{eq:appx_breedbs}
\begin{split}
\hspace{10mm}
    \Qcircuit @C=1.5em @R=2em {
\lstick{}&\qw \bsbal{1} &\qw&\rstick{\ket{\psi}} \qw \\
\lstick{\pbra{m}} &\qw &\qw&\rstick{\ket{\phi}} \qw
%&\qw&\control \qw &\rstick{\ketsub{0}{P_2}} \qw
}\raisebox{-1.5em}{\hspace{8mm}}
\end{split},
\end{equation}
\\
where $\ket{\psi}$ and $\ket{\phi}$ and the input cat states used for breeding. Up to an overall phase, the balanced beamsplitter can be decomposed as~\cite{walshe2020continuous}
\begin{equation}
    \bsop_{12}= e^{-i\op{Q}_1\op{P}_2}\left(\op{S}^\dag_1(\tfrac{1}{2}\ln{2})\op{S}_2(\tfrac{1}{2}\ln{2})\right)e^{i\op{P}_1\op{Q}_2}.
\end{equation}
Performing a homodyne measurement in the $P$-basis to mode two after applying a beamsplitter can thus be written as
\begin{align}
    \brasub{m}{2P}\bsop_{12}&=\brasub{m}{2P}\op{Z}_1^\dag(m)\left(\op{S}^\dag_1(\ln{\small{\sqrt{2}}})\op{S}_2(\tfrac{1}{2}\ln{2})\right)e^{i\op{P}_1\op{Q}_2}\nonumber \\
    &=\op{Z}_1^\dag(m)\op{S}^\dag_1(\ln{\small{\sqrt{2}}})\brasub{m'}{2P}e^{i\op{P}_1\op{Q}_2},
    \label{eq:appx_bsconvert}
\end{align}
where we have used that
\begin{align}
    \op{S}^\dag(r)\ketsub{m}{P}&=\op{Z}(m')\op{S}^\dag(r)\pket{0}\nonumber\\
    &=\pket{m'}
\end{align}
with $m'=2^{-1/2}m$. Finally, we can arrive at the $\opCZ$ gate we want by rewriting
\begin{equation}
\label{eq:appx_cx2cz}
    e^{i\op{P}_1\op{Q}_2}=\op{R}_1(\tfrac{\pi}{2})\op{C}_{Z_{12}}\op{R}_1^\dag(\tfrac{\pi}{2}).
\end{equation}
With the help of Eqs.~\ref{eq:appx_bsconvert} and~\ref{eq:appx_cx2cz}, the starting circuit in Eq.~\ref{eq:appx_breedbs} can be converted to  
\begin{equation}
\begin{split}
\hspace{8mm}
    \Qcircuit @C=1em @R=1.5em {
\lstick{}&\gate{Z^\dag(m)}&\gate{S^\dag(\tfrac{1}{2}\ln{2})}&\gate{R(\tfrac{\pi}{2})}& \ctrl{1} &\rstick{\ket{\psi'}} \qw \\
\lstick{\pbra{m}} &\qw &\qw& \qw & \control \qw& \rstick{\ket{\phi}} \qw
%&\qw&\control \qw &\rstick{\ketsub{0}{P_2}} \qw
}\raisebox{-1.5em}{\hspace{8mm}}
\end{split},
\end{equation}
where $\psi$ has been Fourier transformed to $\ket{\psi'}=\op{R}(\tfrac{\pi}{2})\ket{\psi}$. Thus, if the input cat states are properly rotated, breeding with the beamsplitter is exactly equivalent to breeding with a $\opCZ$ gate up to Gaussian operations that can be undone with further cluster-state processing and feed-forward displacements.

\subsection{Macronode $P(m_2)$}
\label{Appx:macro_Hprob}
The probability of the $Q$-quadrature homodyne measurement $m_2$ in the macronode circuit from Eq.~\ref{eq:macro_dict} in the main text is given by
\begin{equation}
    P(m_2)\propto \int_{-\infty}^\infty \!dy\Big|\brasub{y}{Q_1}\brasub{m_2}{Q_2}\op{B}_{12}e^{-\op{Q}_2^2/2s_2^2}\ketsub{\psi}{1}\ketsub{0}{P_2}\Big|^2,
\end{equation}
where here $\op{B}_{12}$ is a balanced beamsplitter, $s_2=e^{\sq_2}$ is the squeezing in the cluster state, and $\ketsub{\psi}{1}$ is the input quantum state. This can be seen more easily by looking at the reduced circuit in Eq.~\ref{eq:macro_reduced_circuit}, where the $m_2$ homodyne measurement can be seen as occurring immediately after the first beamsplitter. Suppose the input state has the form of a weighted cat state
\begin{equation}
    \ketsub{\psi}{1} \propto \left(A\op{X}(\alpha)+B\op{X}^\dag(\alpha)\right)\op{S}(r_1)\ket{0},
\end{equation}
where we take $A^2+B^2=1$ and let $\alpha\geq 0$ without loss of generality. Using this form of $\psi$, we have that
\begin{align}
    \op{B}_{12}&e^{-\op{Q}_2^2/2s_2^2}\ketsub{\psi}{1}\ketsub{0}{P_2}= \nonumber \\
    &e^{-\frac{(\op{Q}_2-\op{Q}_1)^2}{4s_2^2}}\psi_Q(\tfrac{\op{Q_1}+\op{Q_2}}{\sqrt{2}})\ketsub{0}{P_1}\ketsub{0}{P_2}
\end{align}
and so the integrand is, in the $Q$-basis,
\begin{equation}
    \iint\!dxdx'e^{-\frac{(x-x')^2}{4s_2^2}}\psi_Q(\tfrac{x+x'}{\sqrt{2}})\ketsub{x}{Q_1}\brasub{m_2}{Q_2}{x'}\rangle_{Q_2}.
\end{equation}
Thus, up to a normalization,
\begin{align}
    &P(m_2)= \nonumber \\
    &\int^\infty_{-\infty}\!dx\left|e^{-\frac{(x-m_2)^2}{4s_2^2}}\left(Ae^{-\frac{x+m_2-\alpha\sqrt{2}}{4s_1^2}}+B^{-\frac{x+m_2+\alpha\sqrt{2}}{4s_1^2}}\right) \right|^2.
\end{align}
Performing the integral leads to the equations given in the main text.

\bibliographystyle{unsrtnat}
\bibliography{Pfister_new.bib}

\end{document}